\documentclass[%
 reprint,%
 amssymb, amsmath,%
 aip,pop,%
author-year]{revtex4-1}

\usepackage{bm}%
\usepackage[colorlinks=true,linkcolor=blue,citecolor=blue,urlcolor=blue]{hyperref}%
\expandafter\ifx\csname package@font\endcsname\relax\else
 \expandafter\expandafter
 \expandafter\usepackage
 \expandafter\expandafter
 \expandafter{\csname package@font\endcsname}%
\fi

\usepackage{graphicx}%

\begin{document}

\def\araa {Annual Review of Astron and Astrophys}
\def\apjl {Astrophysical Journal, Letters}
\def\apj {Astrophysical Journal}
\def\aap {Astronomy and Astrophysics}
\def\ssr {Space Science Reviews}
\def\solphys{Solar Physics}
\def\jgr{J.~Geophys.~Res}


\title[Beaming electromagnetic instabilities]{Beaming electromagnetic (or heat-flux) instabilities from the interplay 
with the electron temperature anisotropies}

\author{S. M. Shaaban}
\email{shaaban.mohammed@kuleuven.be}
\affiliation{Centre for Mathematical Plasma-Astrophysics, KU Leuven, Celestijnenlaan 200B, 3001 Leuven, Belgium}\affiliation{Theoretical Physics Research Group, Physics Dept., Faculty of Science, Mansoura University, 35516 Mansoura, Egypt}%
\author{M. Lazar}%
 \affiliation{Centre for Mathematical Plasma-Astrophysics, KU Leuven, Celestijnenlaan 200B, 3001
Leuven, Belgium}%
 \affiliation{Institut f\"ur Theoretische Physik, Lehrstuhl IV: Weltraum- und Astrophysik, Ruhr-Universit\"at Bochum, D-44780 Bochum, Germany}%
 \author{P. H. Yoon}%
 \affiliation{Institute for Physical Science and Technology, University of Maryland, College Park, MD 20742, USA}
\affiliation{Korea Astronomy and Space Science Institute, Daejeon 34055, Korea
School of Space Research, Kyung Hee University, Yongin, Gyeonggi 17104, Korea}
 
\author{S. Poedts}%
\affiliation{Centre for Mathematical Plasma-Astrophysics, KU Leuven, Celestijnenlaan 200B, 3001
Leuven, Belgium}%

\date{\today}%
\revised{-}%

\begin{abstract}
	In space plasmas kinetic instabilities are driven by the beaming (drifting) components and/or the temperature
anisotropy of charged particles. The heat-flux instabilities are known in the literature as electromagnetic 
modes destabilized by the electron beams (or strahls) aligned to the interplanetary magnetic field. 
A new kinetic approach is proposed here in order to provide a realistic characterization of heat-flux 
instabilities under the influence of electrons with temperature anisotropy. Numerical analysis is based 
on the kinetic Vlasov-Maxwell theory for two electron counter-streaming (core and beam) populations with 
temperature anisotropies, and stationary, isotropic protons. The main properties of electromagnetic 
heat-flux instabilities are found to be markedly changed by the temperature anisotropy of electron beam 
$A_b = T_\perp / T_\parallel \ne 1$, leading to stimulation of either the 
whistler branch if $A_b > 1$, or the firehose branch for $A_b<1$. For a high temperature anisotropy 
whistlers switch from heat-flux to a standard regime, when their instability is inhibited by the beam.
  
\end{abstract}

\maketitle

\section{Introduction}\label{sec.1}

Collision-poor plasmas from space are highly susceptible to the instabilities driven by the 
kinetic anisotropies of plasma particles. Thus, the electron strahls, or beaming populations, 
which carry the electron heat-flux in the solar wind, are often associated with enhanced 
electromagnetic (EM) fluctuations \citep{Lengyel1996, Lin1998, Lacombe2014} presumably 
attributed to the so-called heat-flux instabilities \citep{Gary1975}. Contrary to a magnetic 
focusing predicted by the theory, the observations show that strahls lose 
intensity and become wider with heliospheric distance \citep{Maksimovic2005,Stverak2009}. 
In the absence of collisions between particles only the self-generated instabilities can 
be responsible for this degradation \citep{Pagel2007, Saito2007, Gary2007, 
Vocks2005}. These evidences explain the increasing interest for the heat-flux instabilities 
\citep{Saeed2017, Saeed2017b, Shaaban2018}, and understanding their role in this context 
implies a detailed examination in conditions specific to solar wind. 

The heat-flux instabilities are highly conditioned by the electron beam, and, depending on 
the relative beam velocity, two distinct branches can be destabilized. Whistlers 
with a right-handed (RH) circular polarization (in direction of the uniform magnetic field)
are excited by a less energetic beam with velocity lower than thermal speed. Growth rates of the 
whistler heat flux instability (WHFI) show a non-uniform variation, increasing and then decreasing 
with increasing the beaming velocity \citep{Gary1985, Shaaban2018}. In the second branch 
the left handed (LH) firehose heat flux instability (FHFI) \citep{Gary1985} is excited by a more
energetic beam, with growth rates increasing  monotonically with increasing the beam velocity 
\citep{Gary1985, Saeed2017b, Shaaban2018}. Recently, \cite{Shaaban2018} have derived the
beam velocity thresholds for each of these two instabilities in the absence of temperature 
anisotropy, and described the intermediary regime of transition, where both heat-flux 
instabilities may co-exist and compete to each other. It has also been 
shown that effective (counter-)beaming anisotropy is reduced by the suprathermal electrons present in space plasmas, which implies stimulation of the unstable whistlers but inhibition of the firehose instability.

Beams or counter-beaming populations of electrons are ubiquitous in space plasmas, e.g., during 
fast winds and coronal mass ejections, and their kinetic implications cannot be isolated from the 
effects of temperature anisotropies, if both these two sources of free energy are present 
\citep{Stverak2008,Vinas2010}. Here we indeed show that all known properties of the heat-flux 
and temperature anisotropy-driven instabilities may be significantly altered by the interplay of 
beaming electrons and their temperature anisotropy, i.e., $T_\perp \ne T_\parallel$. In fact, in 
such a complex (but realistic) scenario we deal with two distinct triggers of the same unstable
modes. The heat-flux instabilities described above may interplay with the common whistler 
instability (WI) driven by anisotropic electrons with $T_\perp > T_\parallel$ \citep{Gary1996,
Stverak2008,Lazar2018}, or the well-known firehose instability (FHI) excited by the electrons 
with an opposite anisotropy $T_\perp < T_\parallel$ \citep{Paesold1999, Stverak2008, Lazar2018}. 
Recent studies have investigated these regimes for low-beta ($\beta \leqslant 0.4$) electrons, 
and found that WI is inhibited by the beam (growth rates decrease with increasing the beam 
velocity), while FHFI is insensitive to a temperature anisotropy $T_\perp > T_\parallel$ 
of the beam \citep{Saeed2017b}. In an attempt to make a reliable distinction between the 
heat-flux and temperature anisotropy instabilities, our present study provides an extended 
comparative analysis, including the solar wind high-beta ($\beta \geqslant 1$) conditions, 
where kinetic instabilities are expected to be more operative. Suprathermal 
populations are not considered in the present analysis, with the express intention to isolate 
and describe only the instabilities resulting from the cumulative effects of electron beams and 
temperature anisotropy.

In Section \ref{sec.2} we describe the distribution models for the electrons and protons, and derive 
the general dispersion relation for the EM modes, which incorporates the instability cumulative
effects of anisotropic electrons. Whistlers are studied in sec.~\ref{sec.3} and 
firehose instability in sec.~\ref{sec.4}, and then in sec.~\ref{sec.TC} we provide a comparative
study of the instability threshold conditions for different regimes, e.g., WHFI, FHFI, WI and 
FHI, as resulting from the interplay of electron beam and temperature anisotropy. 
Section~\ref{sec.6} summarizes the results obtained in this work with discussions and conclusions.

\section{Dispersion relations}\label{sec.2}

We consider a collisionless quasi-neutral electron-proton plasma with two populations of electrons, 
namely, the core (subscript $a=c$) and the beam (subscript $a=b$), counterstreaming in the protons' frame
\begin{align}
f_e\left(v_\perp, v_\parallel\right)=\eta f_c\left(v_\perp, v_\parallel\right)+\delta f_b\left(v_\perp, v_\parallel\right),   \label{1}
\end{align}
where $\eta=n_c/n_e$ and $\delta=1-\eta$ are relative number densities satisfying neutrality of charge 
of the electrons (subscript $e$) and protons (subscript $p$), $n_e = n_c + n_b = n_p$. Each component 
is a drifting bi-Maxwellian 
\begin{equation}
f_{a}\left(v_{\perp }, v_{\parallel }\right)=\frac{\pi^{-3/2}}{\alpha_{a ,\perp }^{2}\alpha_{a ,\parallel }}
\exp \left[-\frac{v_{\perp }^{2}}{\alpha_{a ,\perp }^{2}} -
\frac{\left(v_\parallel -U_a\right)^{2}}{\alpha_{a ,\parallel }^{2}}\right]   \label{2}
\end{equation}
where drifting velocities $U_a$ are directed along the magnetic field and satisfy a zero net current condition 
$n_c U_c~+~n_b U_b=0$. Thermal velocities $\alpha_{a ,\parallel}=~\sqrt{2k_{B}T_{a ,\parallel}/m_{a}}$ and $\alpha_{a
,\perp} = \sqrt{2k_{B}T_{a ,\perp }/m_{a}}$, are defined in terms of the anisotropic temperature components,  
parallel ($T_\parallel$) and perpendicular ($T_\perp$) to the ambient magnetic field $\bm{B}$.
If protons are bi-Maxwellian, the linear dispersion relations describing the parallel electromagnetic 
modes read \citep{Gary1985}  
\begin{align}
&\frac{c^{2}k^{2}}{\omega ^{2}}=1+\sum_{a=e, c, b }\frac{\omega _{p,a }^{2}}{\omega ^{2}}\left[\xi_a 
Z\left( \xi _a^{\pm }\right)+\Lambda_a\left\{ 1+\xi _a^{\pm}Z\left( \xi _a^{\pm }\right) \right\} \right]  \label{3}
\end{align}
where $c$ is the speed of light, $\omega$ is the wave frequency, $k$ is the wave number, $\omega_{p,a }^{2}=4\pi 
n_{a}e^{2}/m_{a}$ is the plasma frequency, $\pm$ distinguish between the circular right-handed (RH) and left-handed (LH) 
polarizations, respectively, $\Lambda_a=A_a-1$, in terms of temperature anisotropy $A_a =T_{a,\perp},/T_{a,\parallel}$,
$\xi_{a }=\left(\omega -kU_a\right)/(k\alpha_{a,\parallel})$, and 
\begin{equation}
Z\left( \xi _a^{\pm }\right) =\frac{1}{\pi ^{1/2}}\int_{-\infty
}^{\infty }\frac{\exp \left( -x^{2}\right) }{x-\xi _a^{\pm }}dt,\ \
\Im \left( \xi _a^{\pm }\right) >0  \label{4}
\end{equation}
is the plasma dispersion function \citep{Fried1961} of argument
\begin{equation*}
\xi _{a }^{\pm }=\frac{\omega \pm \Omega _a-kU_a}{k\alpha_{a,\parallel }}.
\end{equation*}
For isotropic protons, we can rewrite (\ref{3}) 
\begin{align} 
&\frac{\tilde{w}}{\tilde{k} \sqrt{\mu \beta_{p}}} Z\left(\frac{\mu \tilde{w} \pm 1}{\tilde{k}\sqrt{\mu \beta_{p}}}\right)+\eta  \left[ \Lambda_c+\frac{\left(\Lambda_c+1\right)\left(\tilde{w}+u_c \tilde{k}\right)  \mp \Lambda_c}{\tilde{k} \sqrt{\beta_c}}\right.\nonumber \\
&\left. \times Z\left(\frac{\tilde{w} \mp 1+u_c \tilde{k}} {\tilde{k}\sqrt{\beta_{c}}}\right) \right]+\delta  \left[\Lambda_b+\frac{\left(\Lambda_b+1\right) \left(\tilde{w}-u_b \tilde{k}\right) \mp \Lambda_b} {\tilde{k} \sqrt{\beta_b}}\right.\nonumber \\
&\left. \times Z\left(\frac{\tilde{w} \mp 1-u_b \tilde{k}}{\tilde{k}\sqrt{\beta_{b}}}\right) \right]=\tilde{k}^2 \label{eq:dis}
\end{align}
in terms of the normalized quantities, $\tilde{k}=kc/\omega_{p,e}$, $\tilde{w}= \omega/|\Omega_e|$, the 
proton--electron mass ratio $\mu = m_p/m_e$, the plasma beta for the population of sort $a$, 
$\beta_a = 8\pi n_a k_B T_{a, \parallel}/\bm{B}^2$, and relative velocities of the beam and core 
components, $u_b= U_b\, \omega_{p,e}/(c\, \Omega_e)$ and $u_c=\delta\, u_b/(1-\delta)$, respectively. 

\begin{table}[b]
\centering
\caption{Parameters for the $j$-component of electrons\label{t1}}
\begin{tabular}{c c c c c c c} 
\hline
& Beam electrons ($h$) & & Core electrons ($c$) && Ions ($i$) &  \\ 
\hline 
$n_j/n_i$ & 0.05 && 0.95  && 1.0  \\ 
$T_{j,\parallel}/T_{i,\parallel}$ & 10.0 && 1.0 && 1.0 \\
$m_j/m_i$ & 1/1836 && 1/1836 && 1.0 \\
$T_{j, \perp}/T_{j,\parallel}$ & $\neq$ 1.0 && $\neq$ 1.0 && 1.0 \\
\hline
\end{tabular}
\end{table}
%

Plasma parameters used in the numerical calculations are given in Table \ref{t1}, unless otherwise 
specified. These parameters are inspired from the solar wind observations providing
electron data from different heliocentric distances. Relevant are the electron data making 
distinction between core and beaming (strahl) components, see the density contrasts in \cite{Stverak2009} 
(Figs.~4 and 8), which suggest an average (representative) value $\delta = n_b/n_0 = 0.05$. 
Presuming that strahl and halo electrons comprise the main source of the heat flux transport in 
the solar wind, for the beam-core temperature contrast the observations estimate a variation 
between 3 and 13 along the heliocentric distance with an increasing tendency during the fast winds
\citep{Pilipp1987, Vinas2010, Pierrard2016}. In this case we assume $T_{b,\parallel}/T_{c,\parallel}
=10$. Seeking generalization, for the temperature anisotropy, which is a key parameter in this study, 
we adopt moderate values typically reported in the solar wind \citep{Pilipp1987, Phillips1989, Stverak2008, 
Pierrard2016}. For the beam velocity values are chosen to ensure conditions for both the WHFI and
FHFI, which may also be relevant for the beaming electrons in space plasmas \citep{Pulupa2014}.

\begin{figure}[t]
\centering 
\includegraphics[width=17pc]{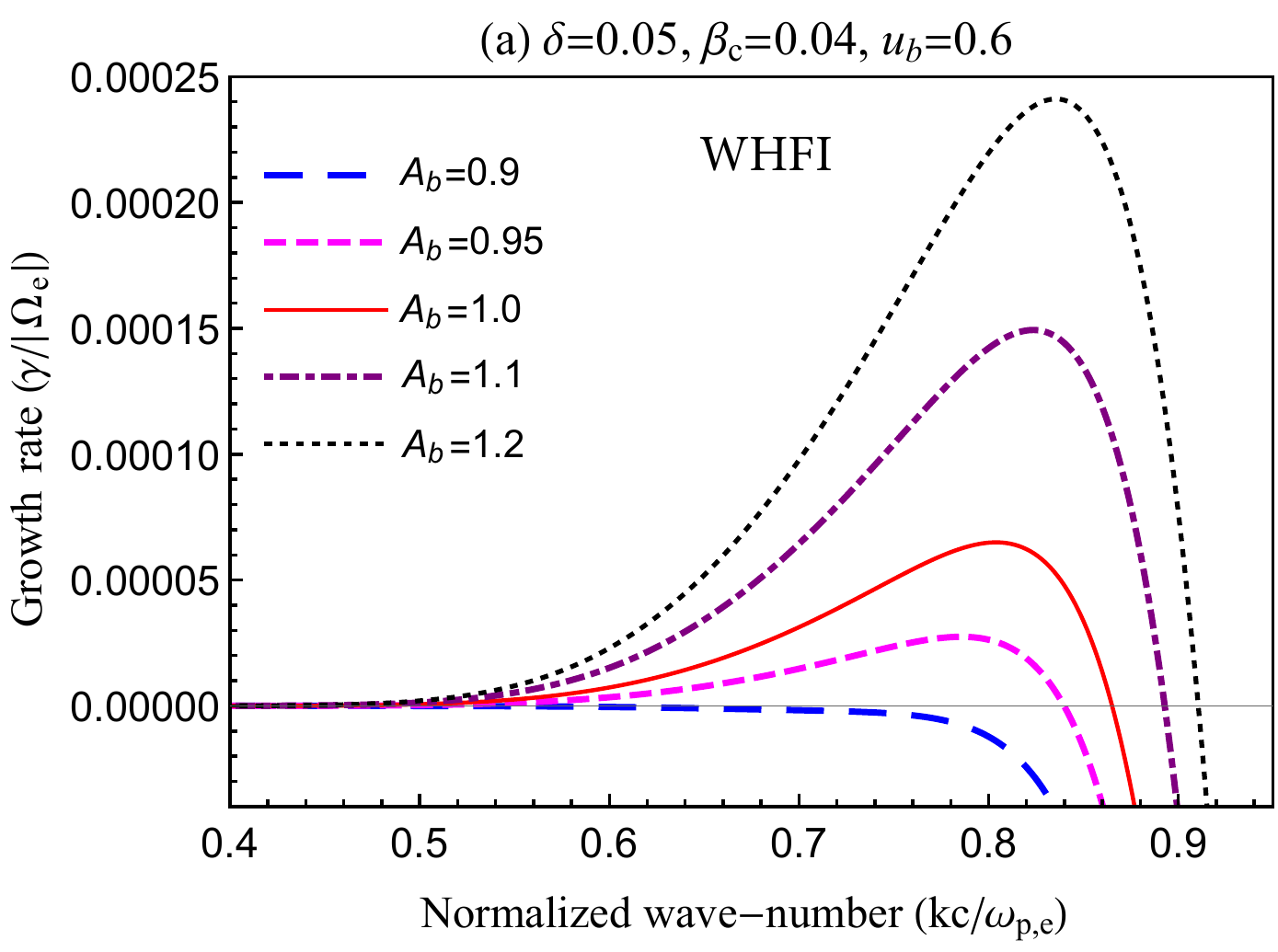}
 \includegraphics[width=16pc]{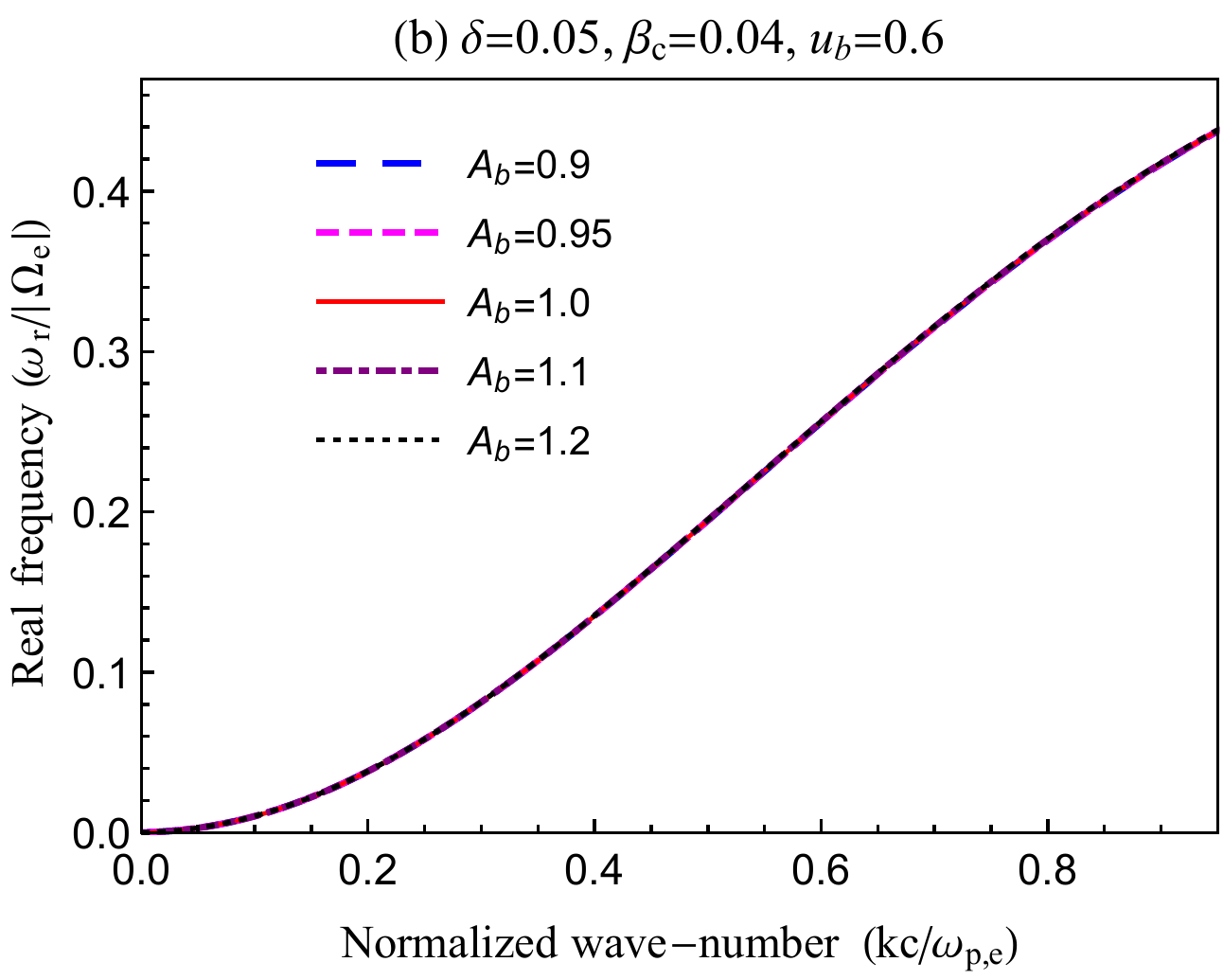}
\caption{WHFI: Effects of the beam anisotropy $A_b$ on the growth rates (panel a) and wave frequencies 
(panel b). The plasma parameters are mentioned in each panel velocity.}
\label{f1}
\end{figure}
%
\begin{figure}[t]
\centering 
\includegraphics[width=17pc]{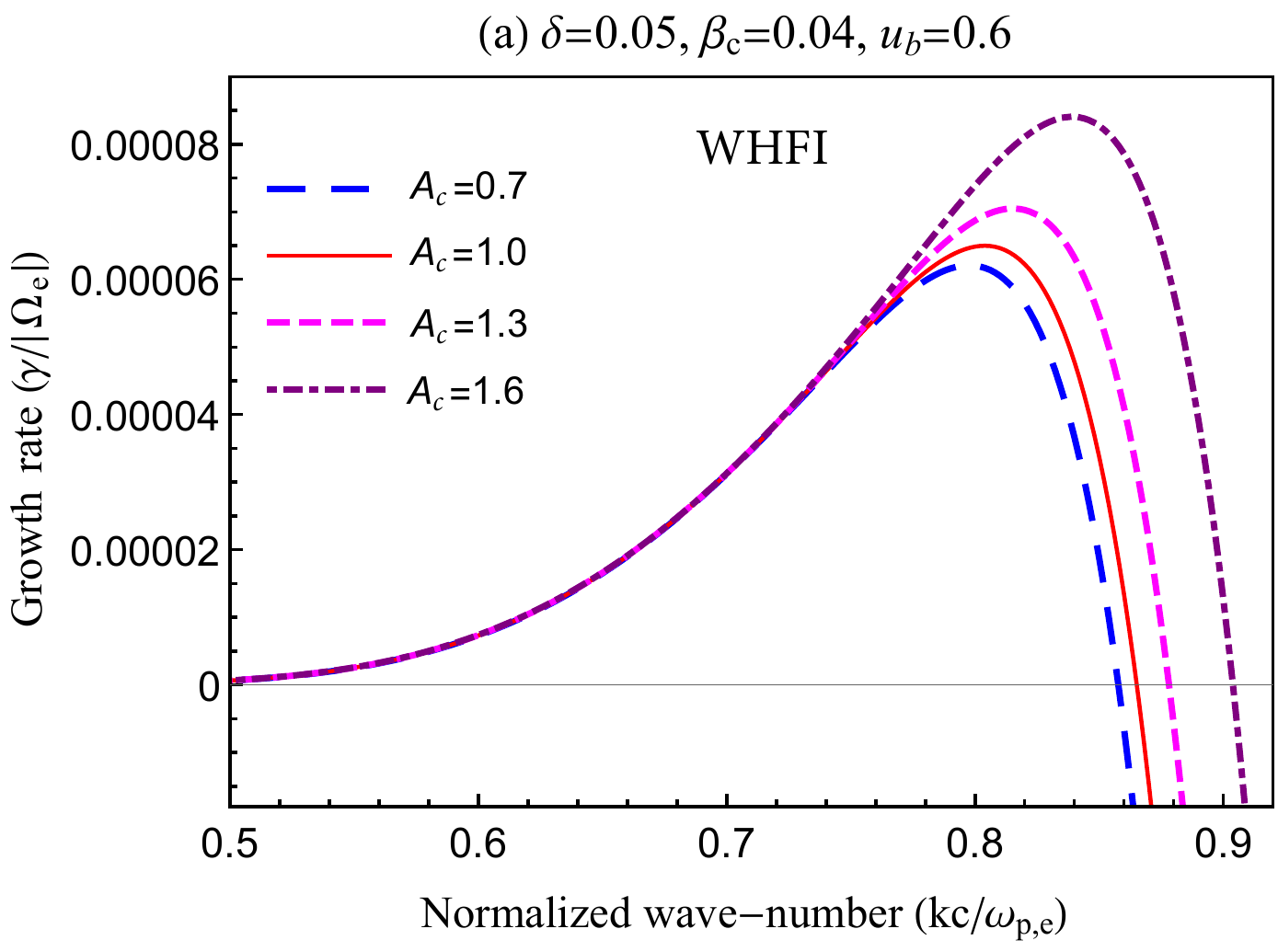}
 \includegraphics[width=16pc]{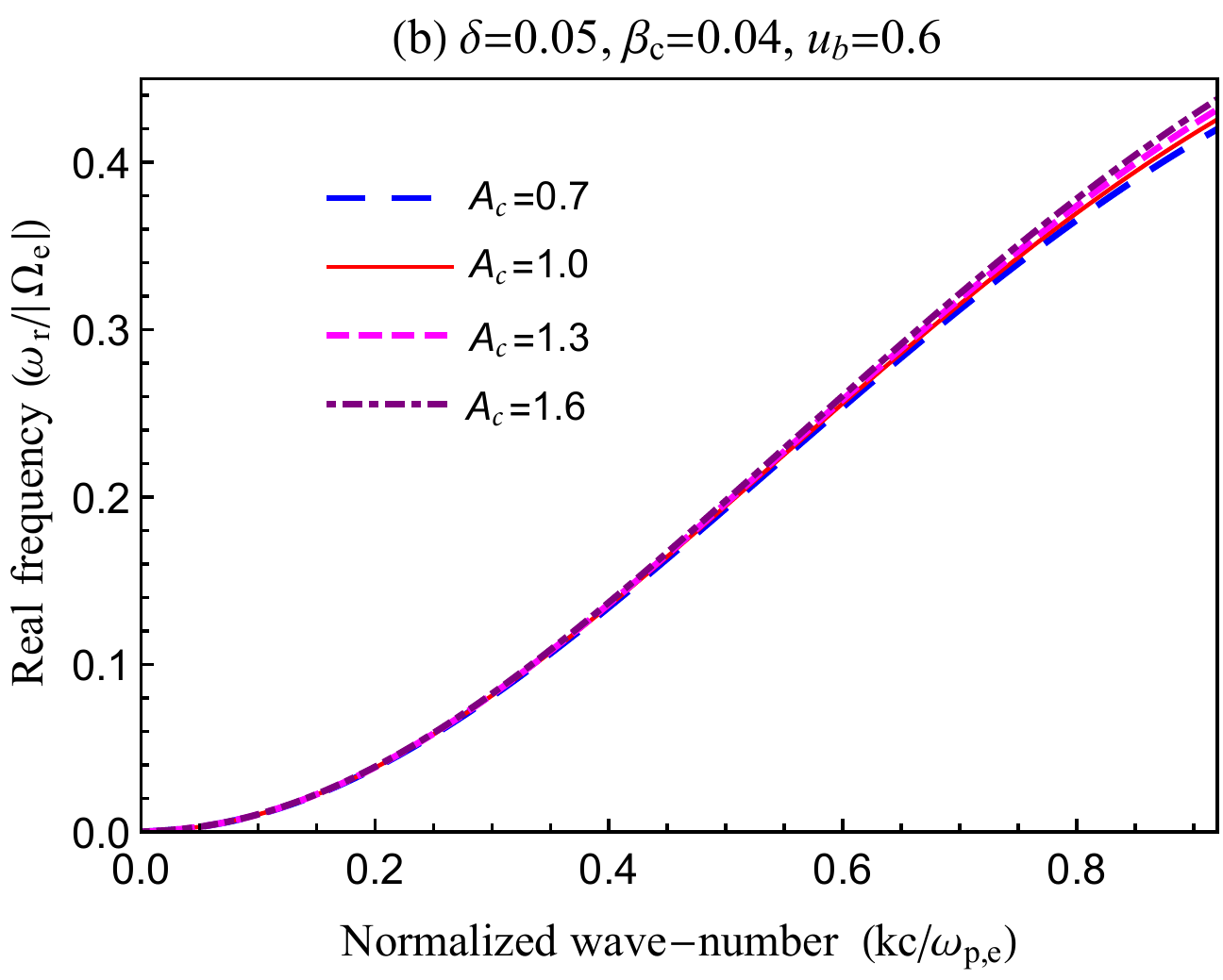}
\caption{WHFI: Effects of the core anisotropy $A_c$ on the growth rates (panel a) and wave frequencies (panel b). 
The plasma parameters are mentioned in each panel.}
\label{f2}
\end{figure}

\begin{figure*}[t]
\centering 
\includegraphics[width=17pc]{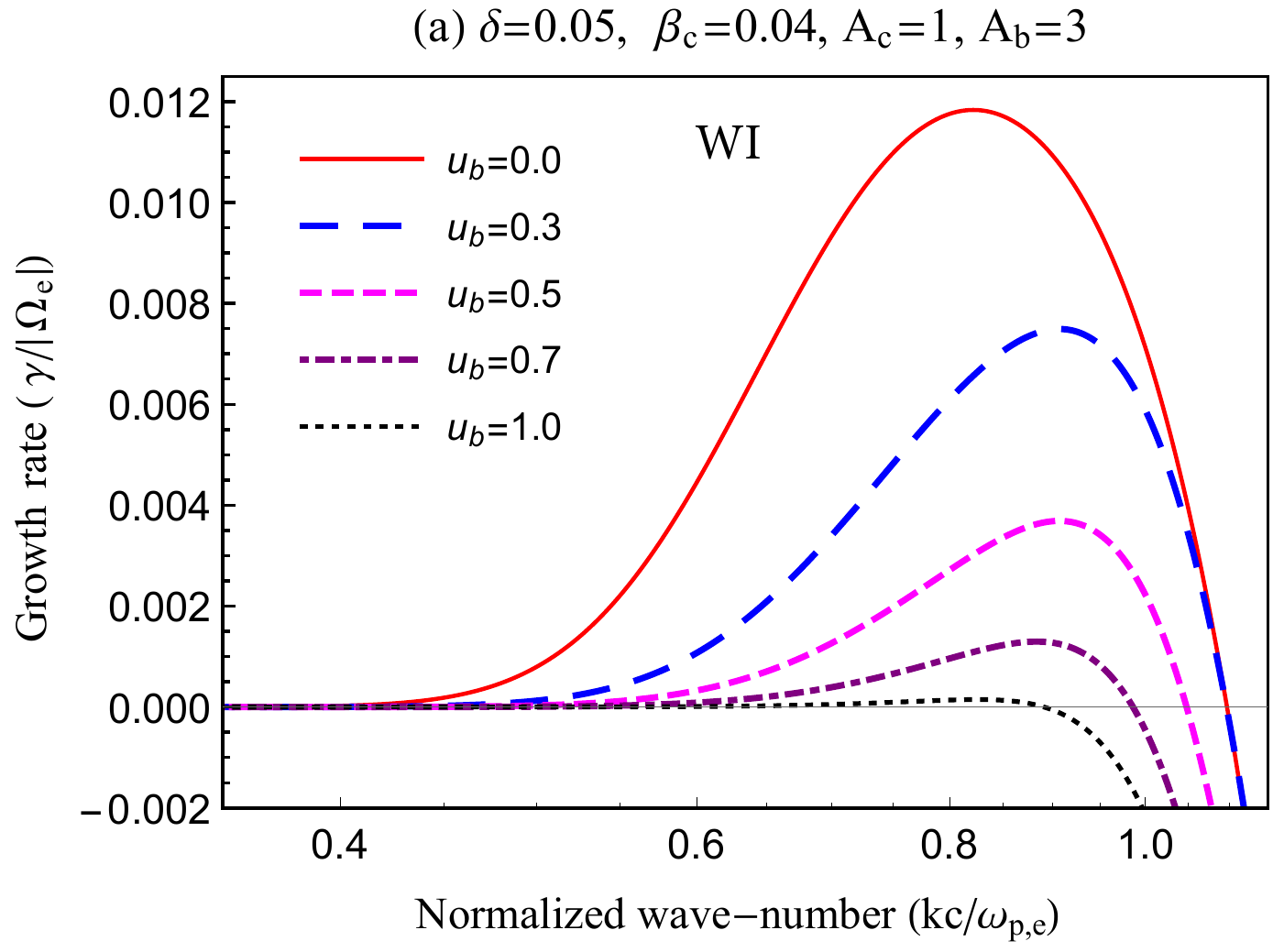} 
\includegraphics[width=17pc]{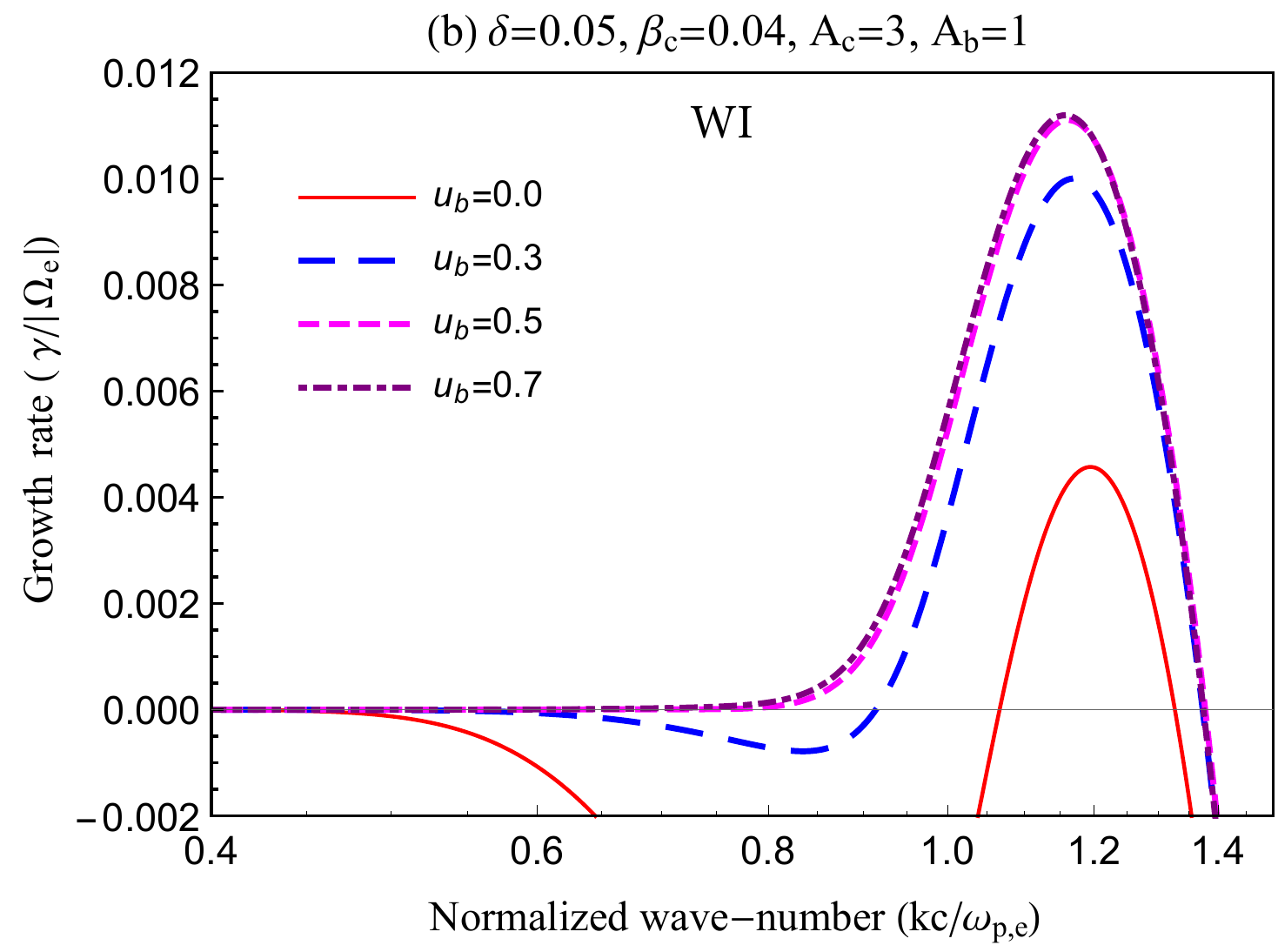}
\caption{WI: Effects of the beam velocity $u_b$ on the growth rates of the WI driven either by an anisotropic beam 
$A_b=3.0$ (panel a) or by an anisotropic core $A_c=3.0$ (panel b). The other plasma parameters are mentioned in each panel. }
\label{f3}
\end{figure*}

\begin{figure*}[t]
\centering 
\includegraphics[width=17pc]{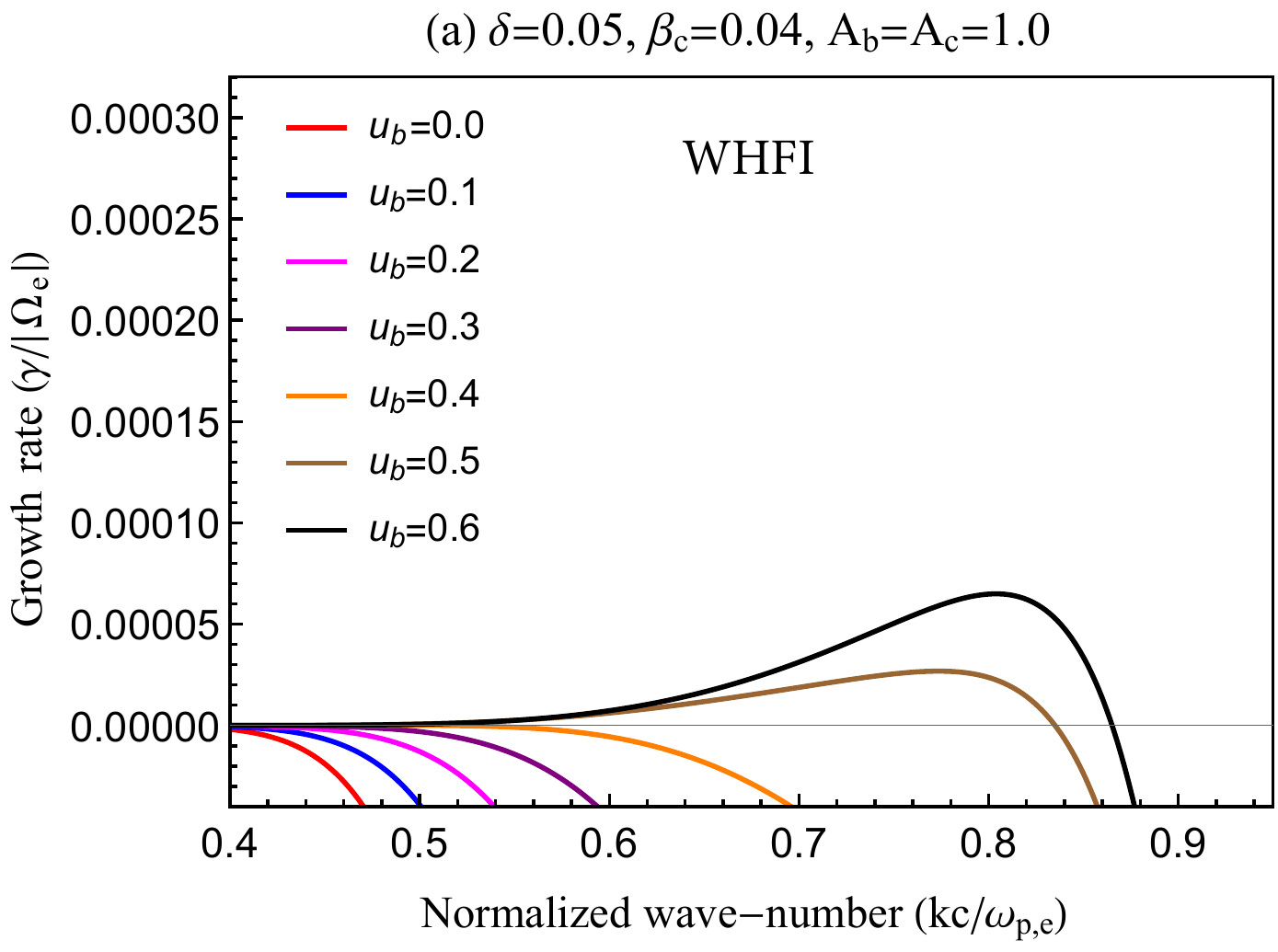} ~~~\includegraphics[width=17pc]{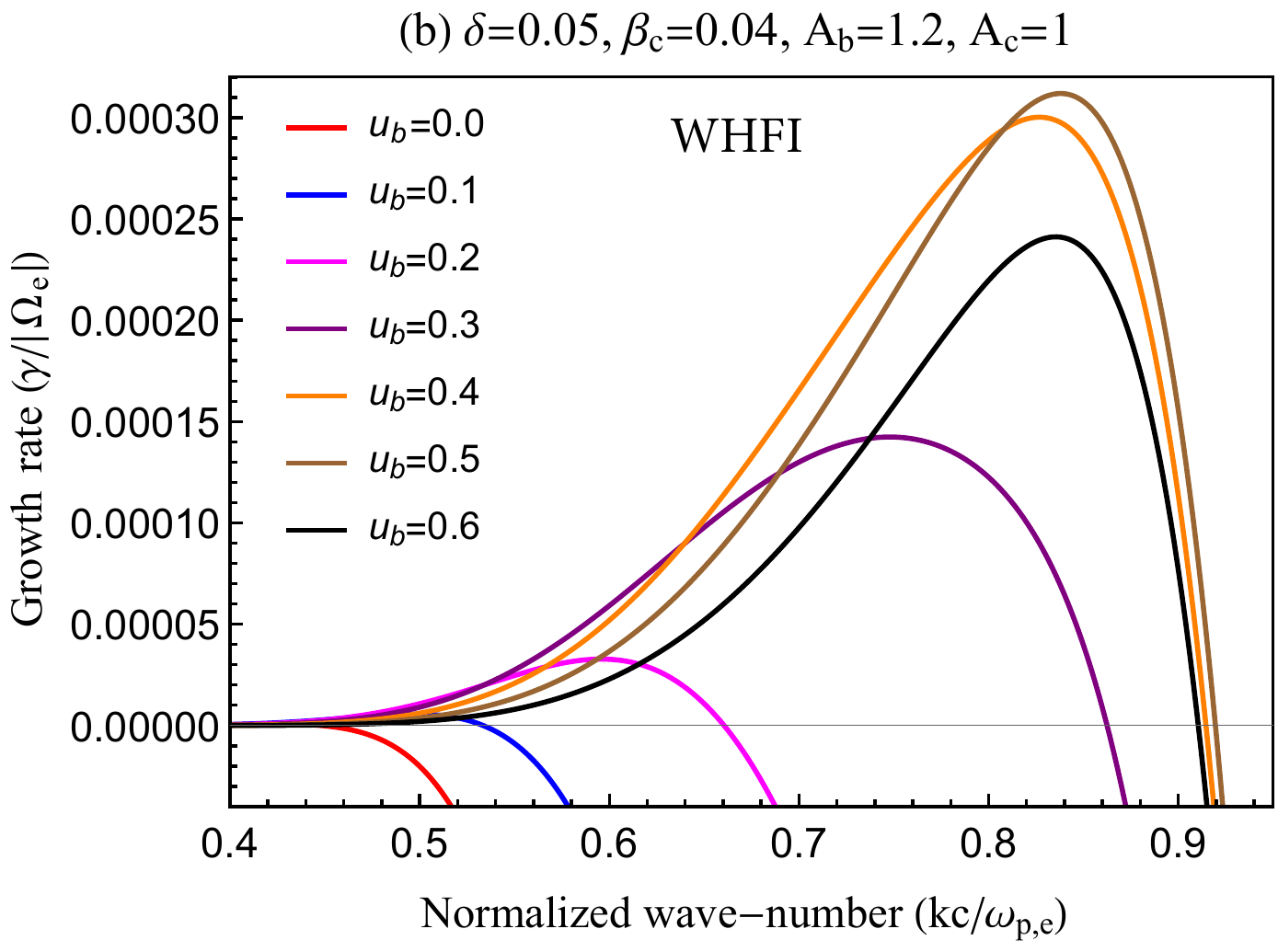}
\includegraphics[width=17pc]{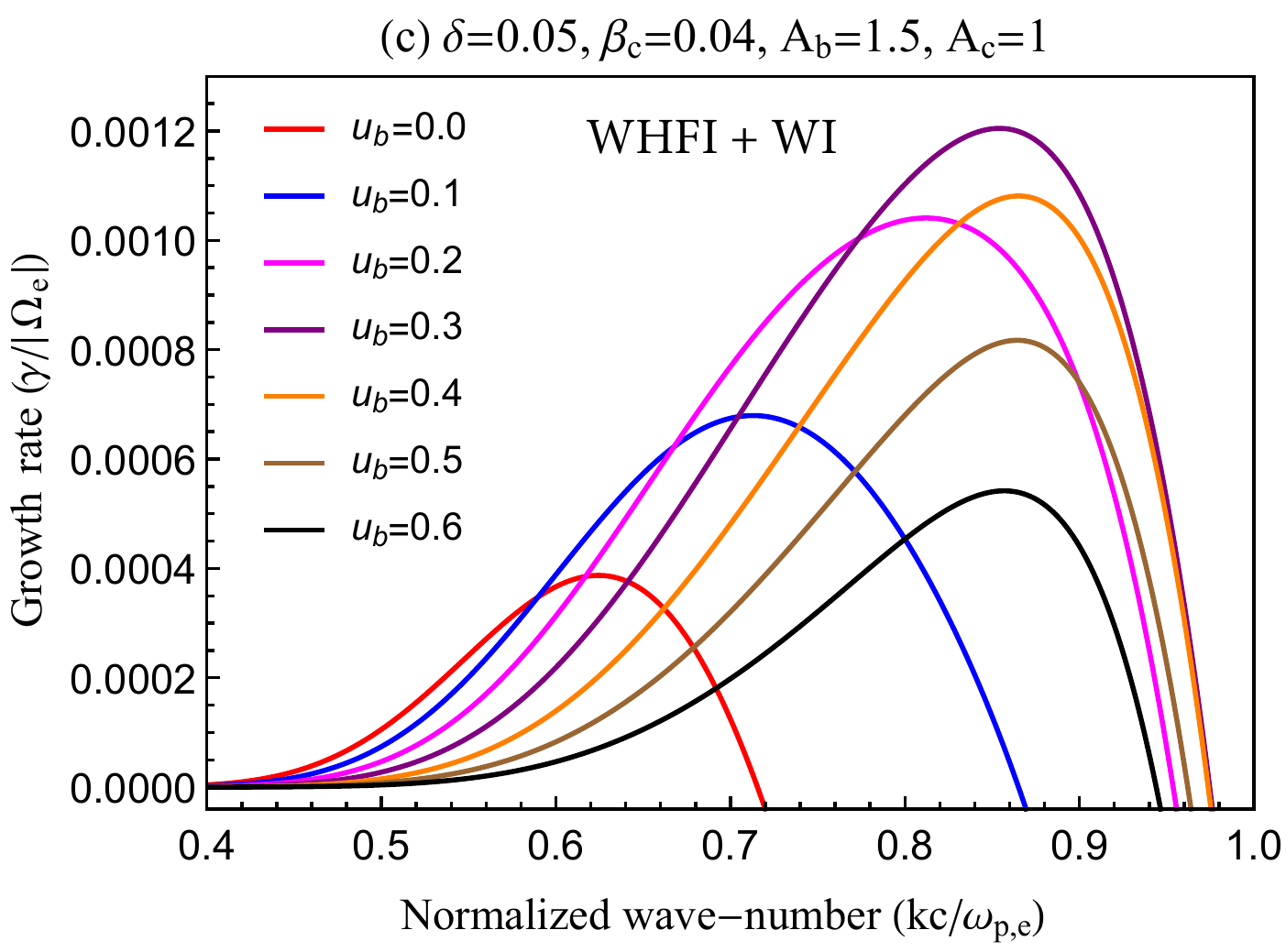} ~~~\includegraphics[width=17pc]{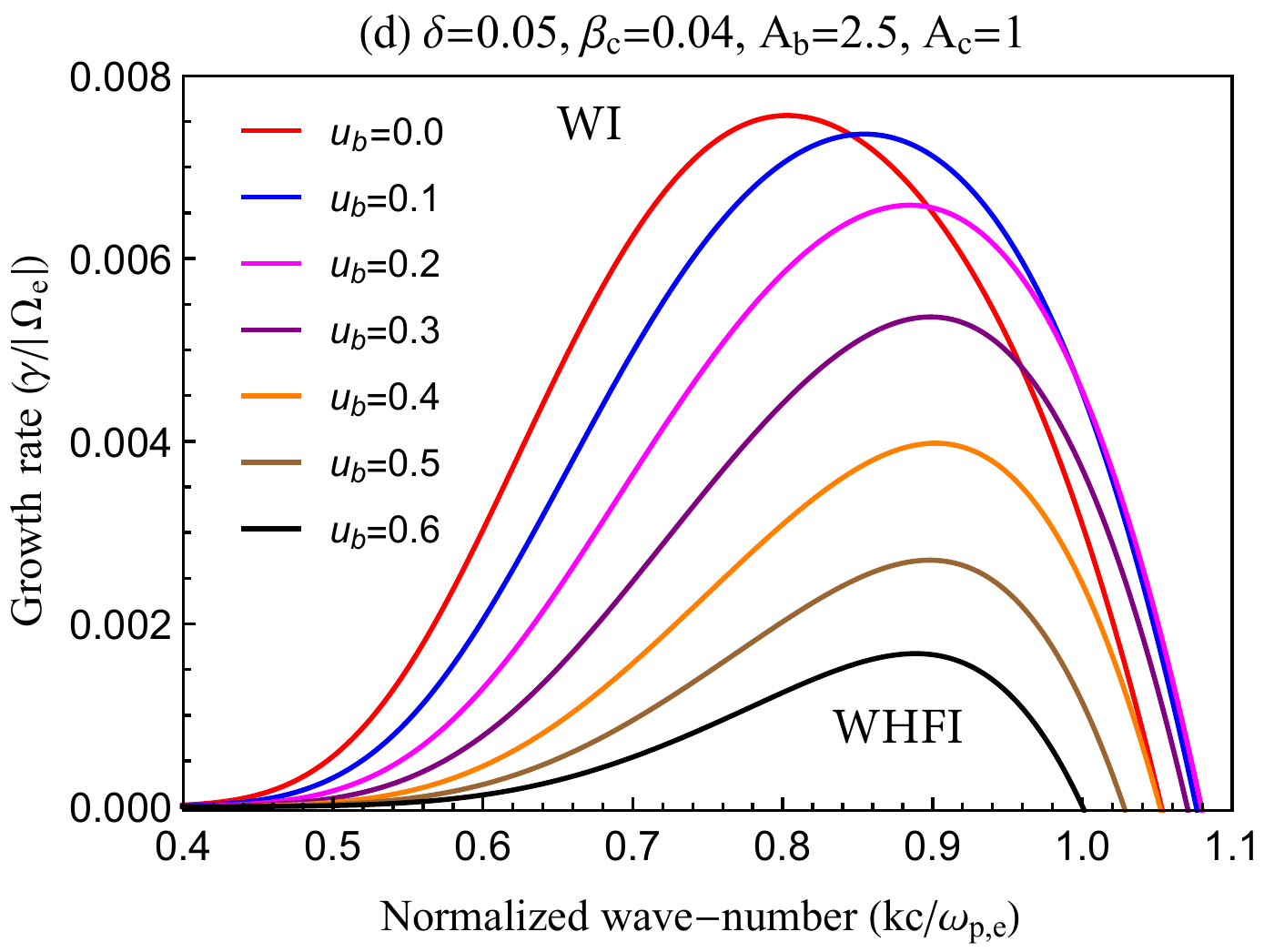}
\caption{Effect of the of the beam velocity $u_b$ on the cumulative whistler instabilities for beam anisotropies $A_b=1.0 $ (panel a), $A_b=1.2$ (panel b), $A_b=1.5 $ (panel c), $A_b=2.5$ (panel d). The plasma parameters are mentioned in each panel. }
\label{f4}
\end{figure*}
%

\section{Unstable whistler modes}\label{sec.3}

We start the analysis with the dispersive characteristics of the whistler modes driven unstable by 
the interplay of the beam-core counter-streaming electrons and their temperature anisotropies. 
These are solutions of the dispersion relation \eqref{3} for the RH modes with $\xi_p^+$.
The less energetic beams are susceptible to the whistler heat flux instability (WHFI) \citep{Shaaban2018},
which is examined in Figures~\ref{f1} and \ref{f2} for the following plasma parameters $\delta=0.05$, 
$\beta_c=\beta_p=0.04$, $u_b=0.6$. In Figure~\ref{f1} we isolate the effects 
of the beam anisotropy by considering isotropic core with $A_c=1.0$, and show the influence of the 
beam anisotropy $A_b=0.9, 0.95, 1.0, 1.1, 1.2$ on the growth rates (panel a) and wave-frequencies (panel 
b) of WHFI. Growth rates are markedly enhanced by increasing the temperature anisotropy in perpendicular direction, 
$A_b>1$, and are inhibited by an opposite anisotropy in parallel direction, $A_b<1$. The corresponding 
wave-frequencies remain unaffected by the variation of the beam temperature anisotropy. 
These unstable solutions are derived for relatively low anisotropies of the beam ($0.9\leqslant A_b \leqslant 1.2$), 
and  a low plasma beta of the core $\beta_c=0.04$, to avoid the whistler or firehose instability effects driven by the temperature anisotropies.
For higher anisotropies of the beam $A_b>1.2$ whistlers exhibit significant growth rates characteristic 
to the whistler instability (WI) driven by the temperature anisotropy, and it becomes difficult to 
distinguish between the WHF and WI regimes, as discussed later in Figures~\ref{f4}. 
In Figure~\ref{f2} we assume an isotropic beam ($A_b=1$) and outline the effect of the core anisotropy $A_c=0.7, 
1.0, 1.3, 1.6$ on the growth rates (panel a) and wave-frequencies (panel b) of WHFI. The growth rates change only
slightly, being enhanced by the core anisotropy in perpendicular direction $A_c > 1$, but inhibited by an 
opposite anisotropy in parallel direction $A_c < 1$. Clearly, the anisotropic beam has a higher influence on the
instability: for $A_b=1.1$ maximum growth rate in Figure~\ref{f1} is three times higher than that obtained for 
$A_c=1.6$ in Figure~\ref{f2}. For reference, the growth rates for isotropic isotropic temperatures $A_b = A_c=1$, 
are displayed in both figures with red solid lines. 

\begin{figure}[t]
\centering 
\includegraphics[width=17pc]{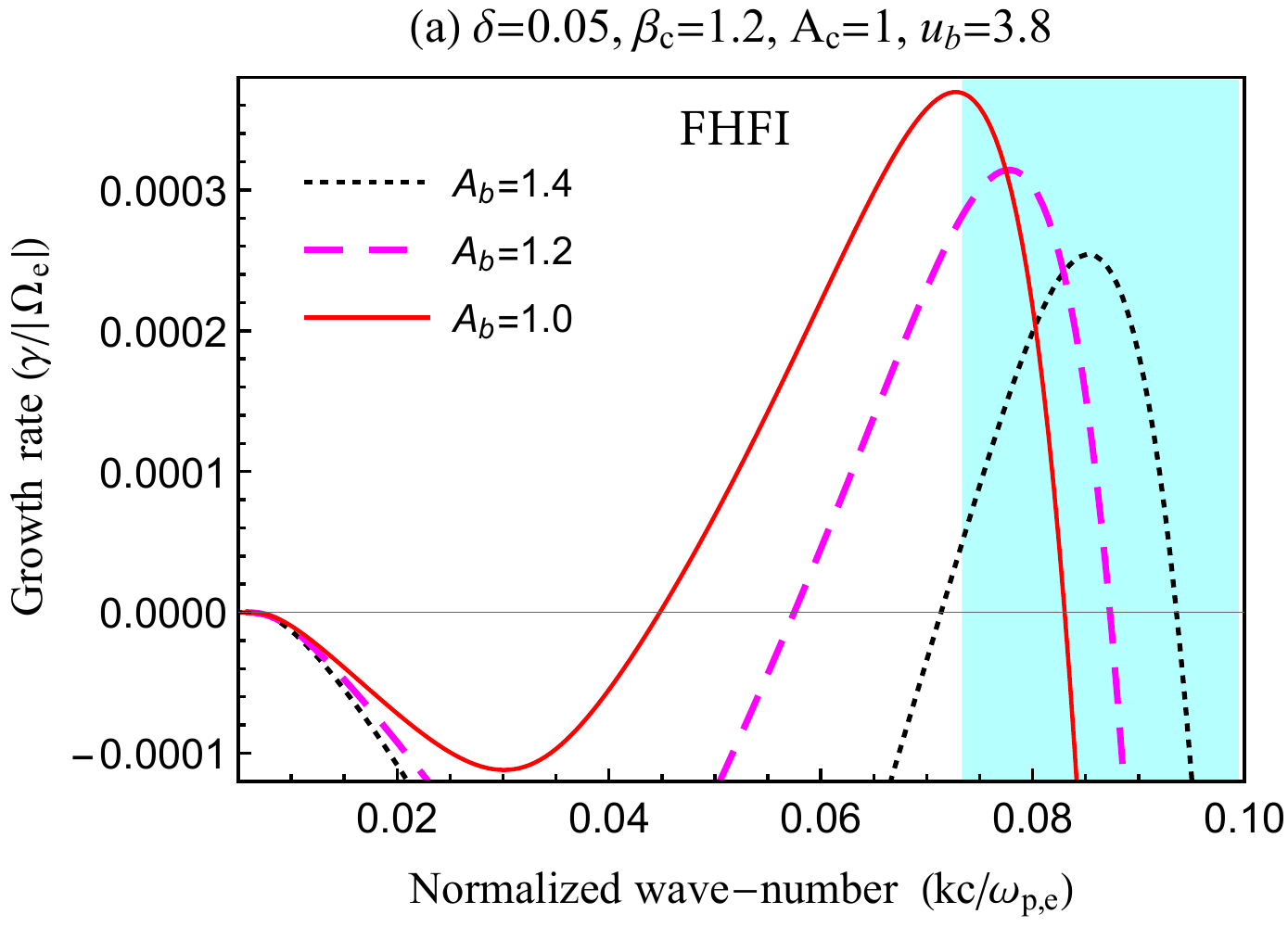}
 \includegraphics[width=16.5pc]{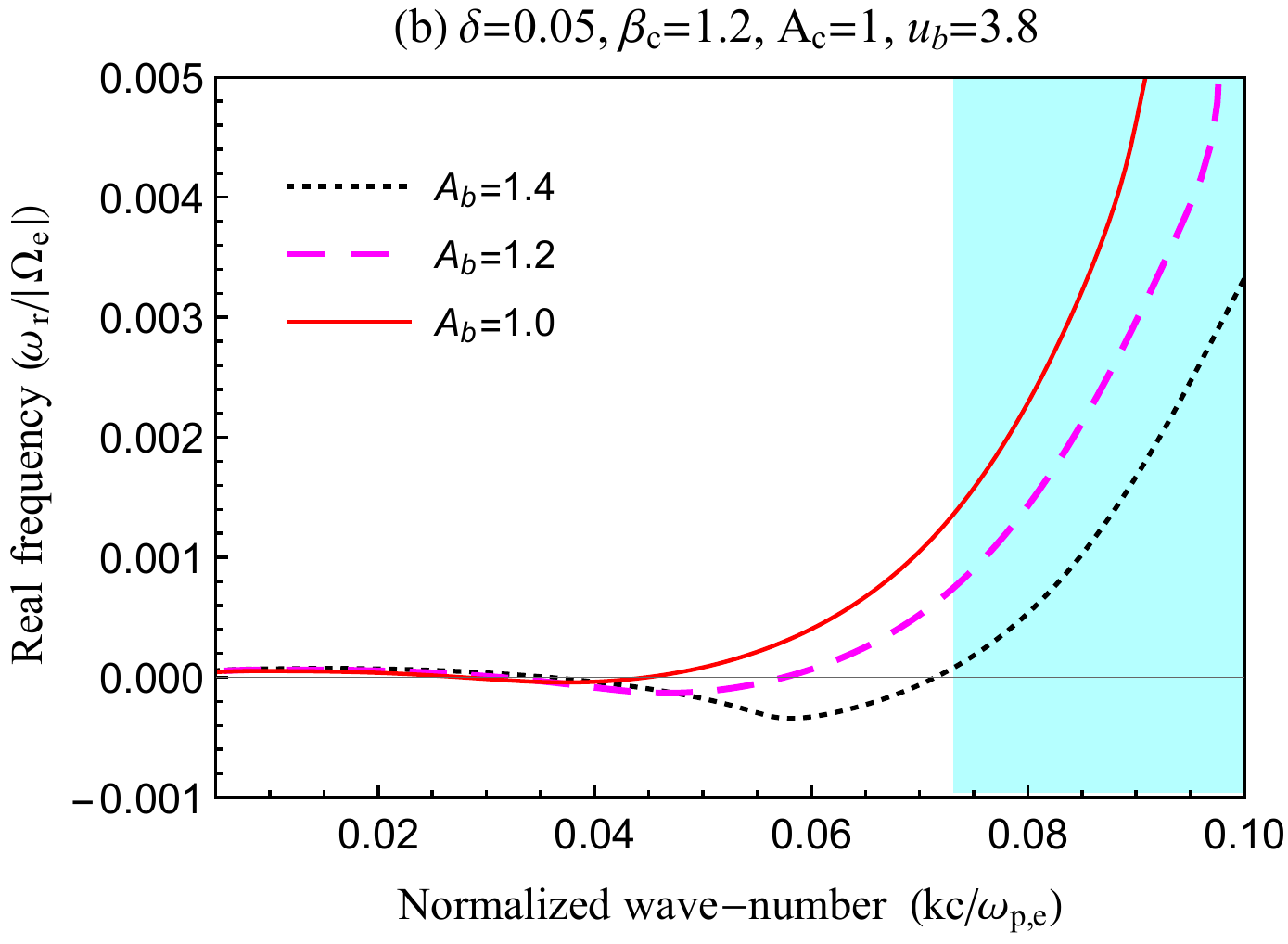}
\caption{FHF: Effect of beam anisotropy $A_b > 1 $ on the growth rates (panel a) and wave frequency (panel b).} \label{f5}
\end{figure}
\begin{figure}[t]
\centering 
\includegraphics[width=16.5pc]{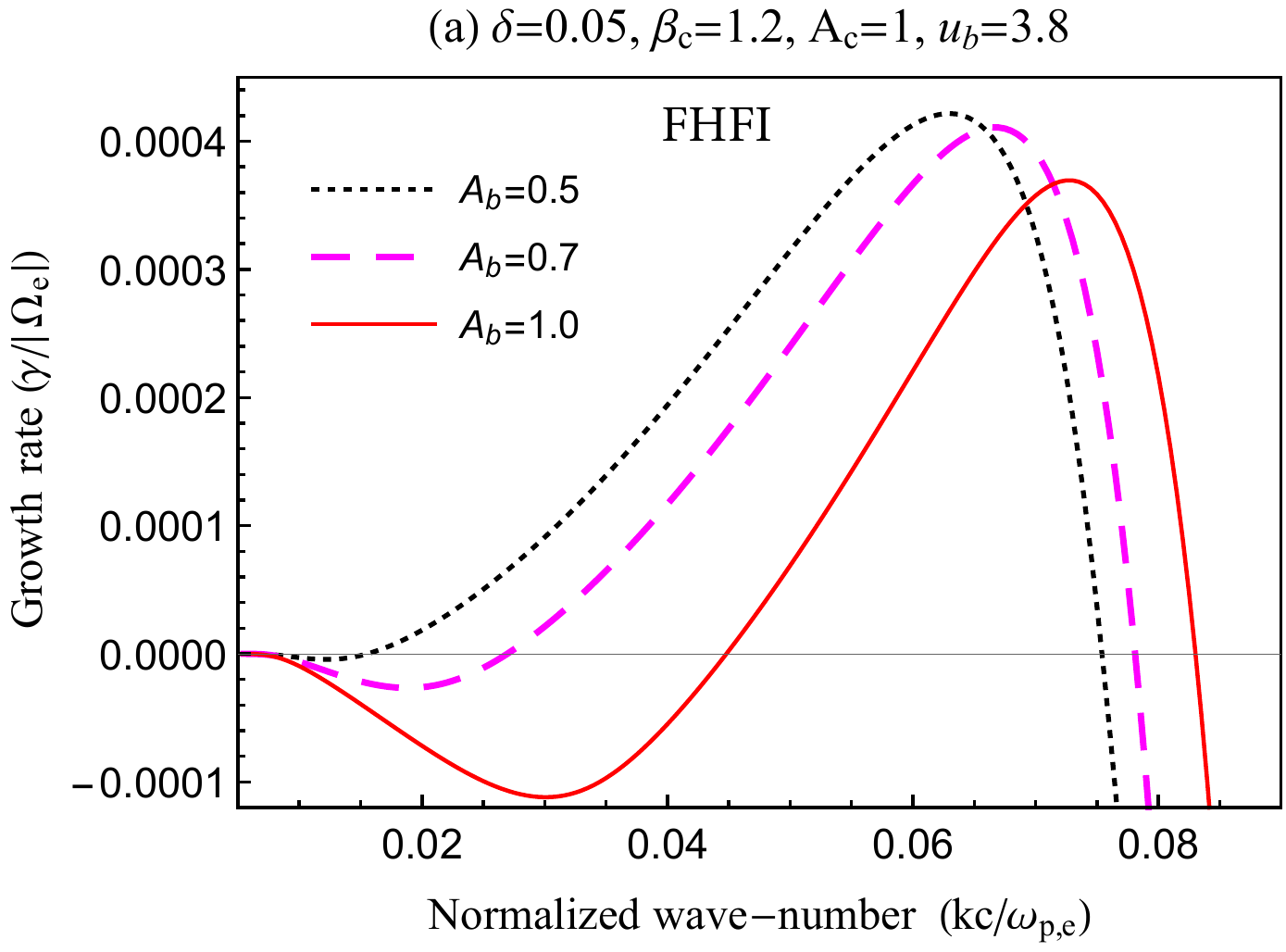}
\includegraphics[width=16pc]{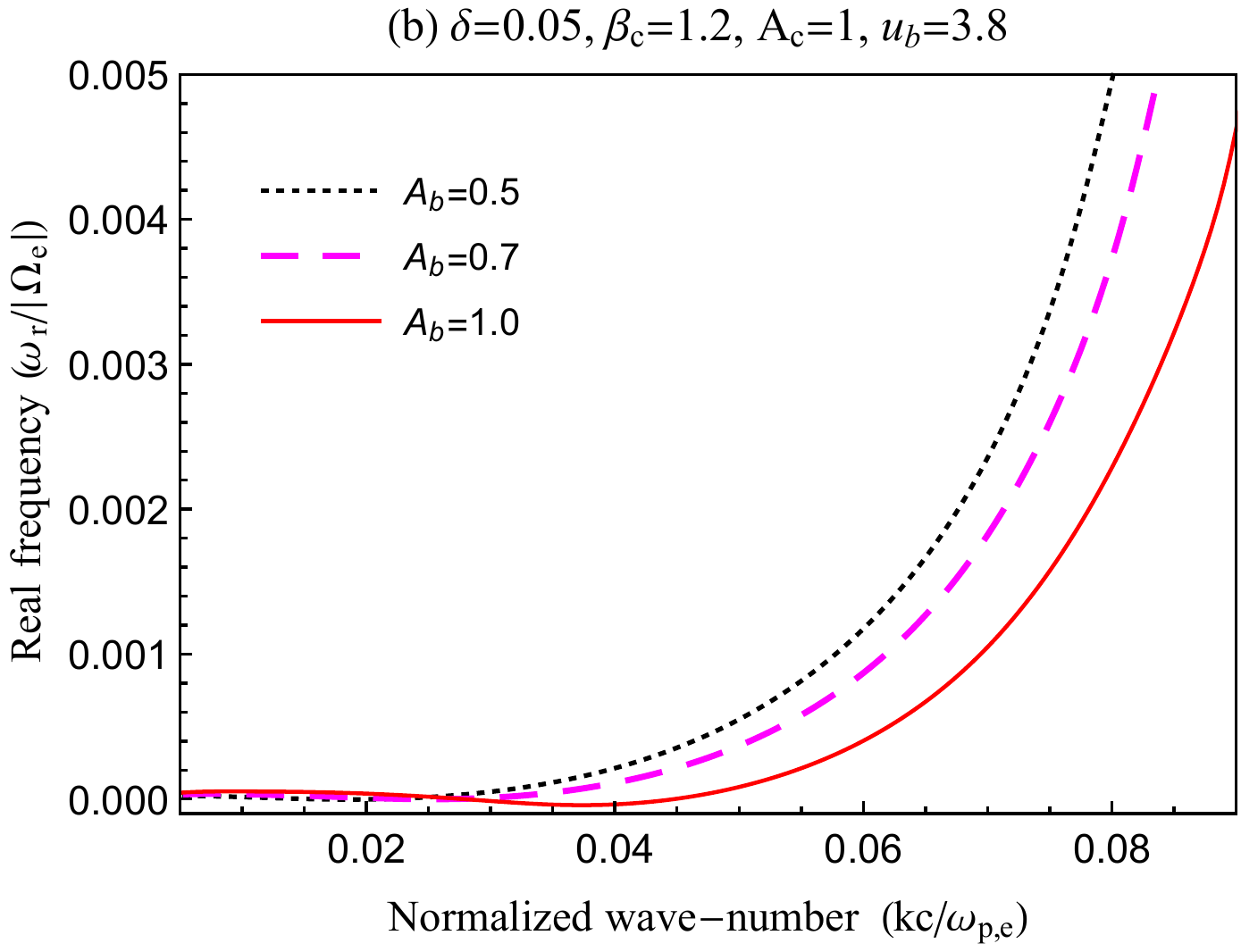}
\caption{FHF: The same in Figure~\ref{f6} but for $A_b<1.0$.} \label{f6}
\end{figure}

The temperature anisotropy driven instabilities are usually studied in the absence of beaming components.
Figure~\ref{f3} shows the effect of beaming velocity $u_b$ on the WI driven by a higher temperature 
anisotropy, for $\delta=0.05$ and $\beta_c=\beta_p=0.04$. When WI is driven by anisotropic beam with 
$A_b=3.0$ (panel a), the effective anisotropy diminishes with increasing the beam speed $u_b$ and the 
instability is inhibited, reducing growth rates and the intervals of unstable wave-numbers. By contrast, 
growth rates driven by the anisotropic core with $A_c =3.0$ (panel b) are enhanced by increasing $u_b$, 
and saturate for $u_b\geqslant 0.5$, resembling a regime characteristic to WHFI. The wave frequency (not 
shown here) only slightly decreases by increasing $u_b$. WHFI has dispersive characteristics similar to 
WI. Both instabilities are driven by resonant electrons and display maximum growth rates in directions 
parallel to the background magnetic field (when the modes are right-hand circularly polarized) \citep{Gary1993, 
Lazar2018}. However, WHFI and WI represent two distinct regimes of whistler modes, destabilized by, 
respectively, the beam $u_b$ and temperature anisotropy $A_b>1$.

\begin{figure*}[t]
\centering 
\includegraphics[width=17.5pc, trim={0 0 14.8cm 0},clip]{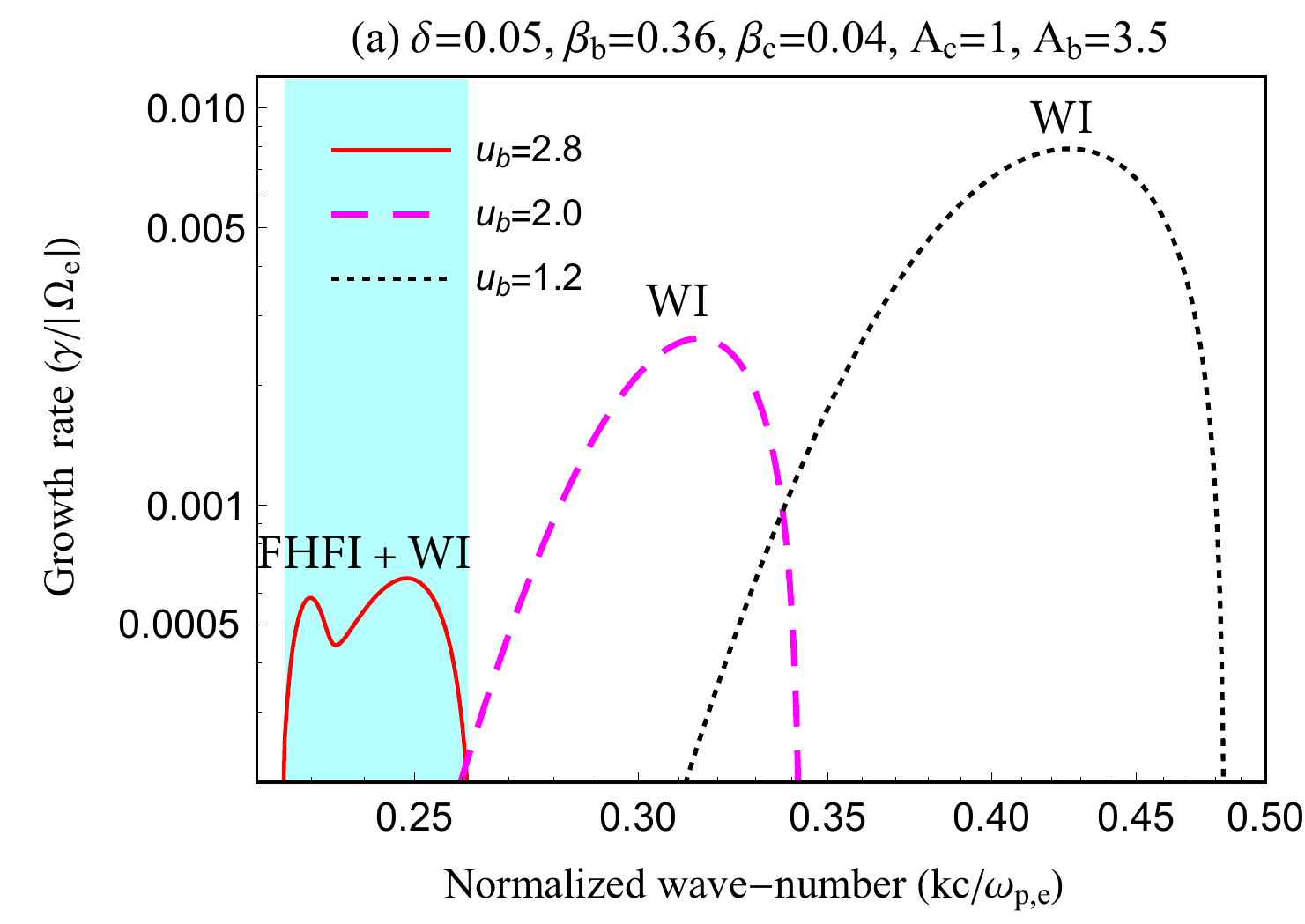} ~~~\includegraphics[width=17pc, 
trim={15.5cm 0 0 0},clip]{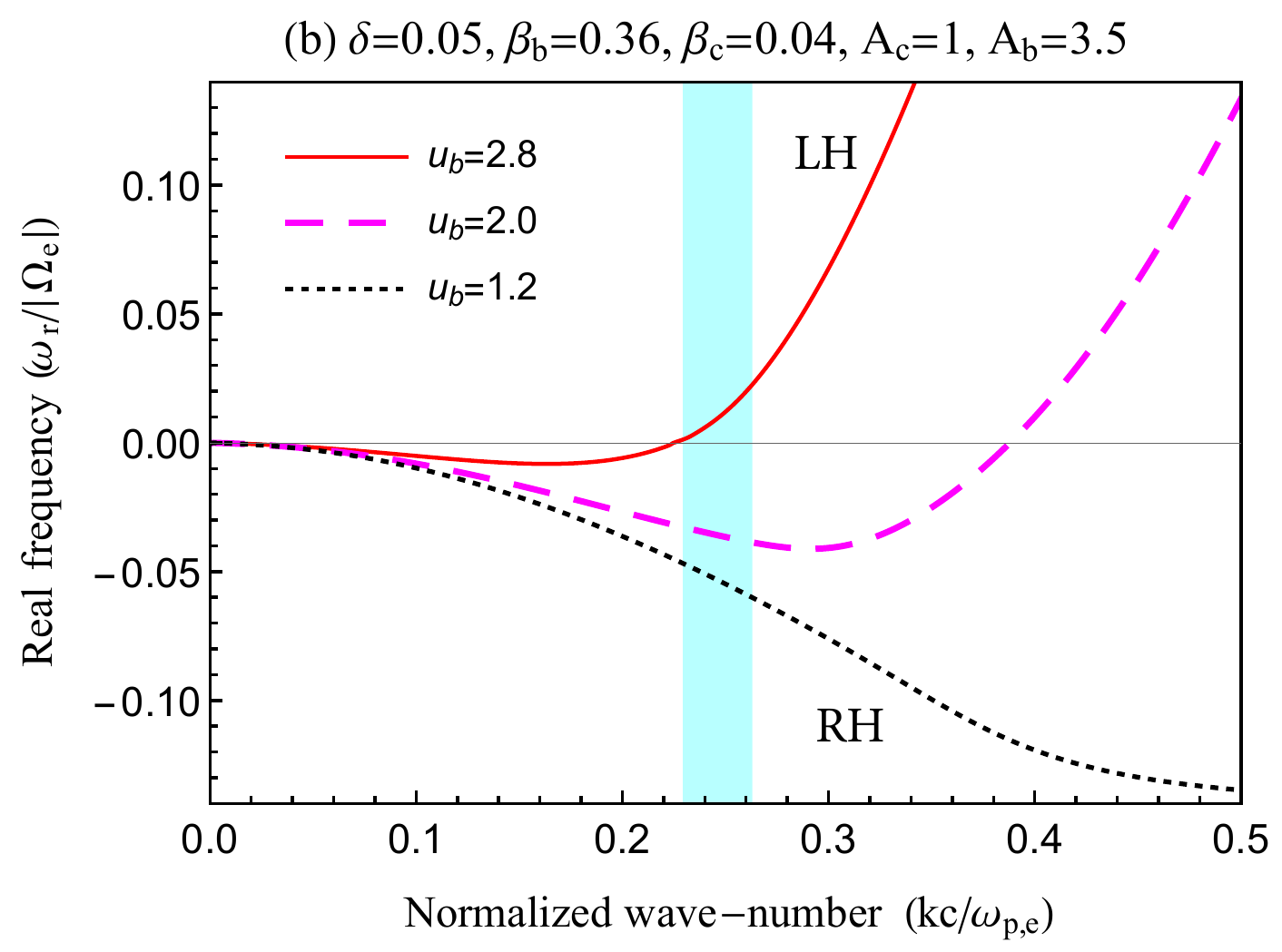}
\includegraphics[width=16.5pc]{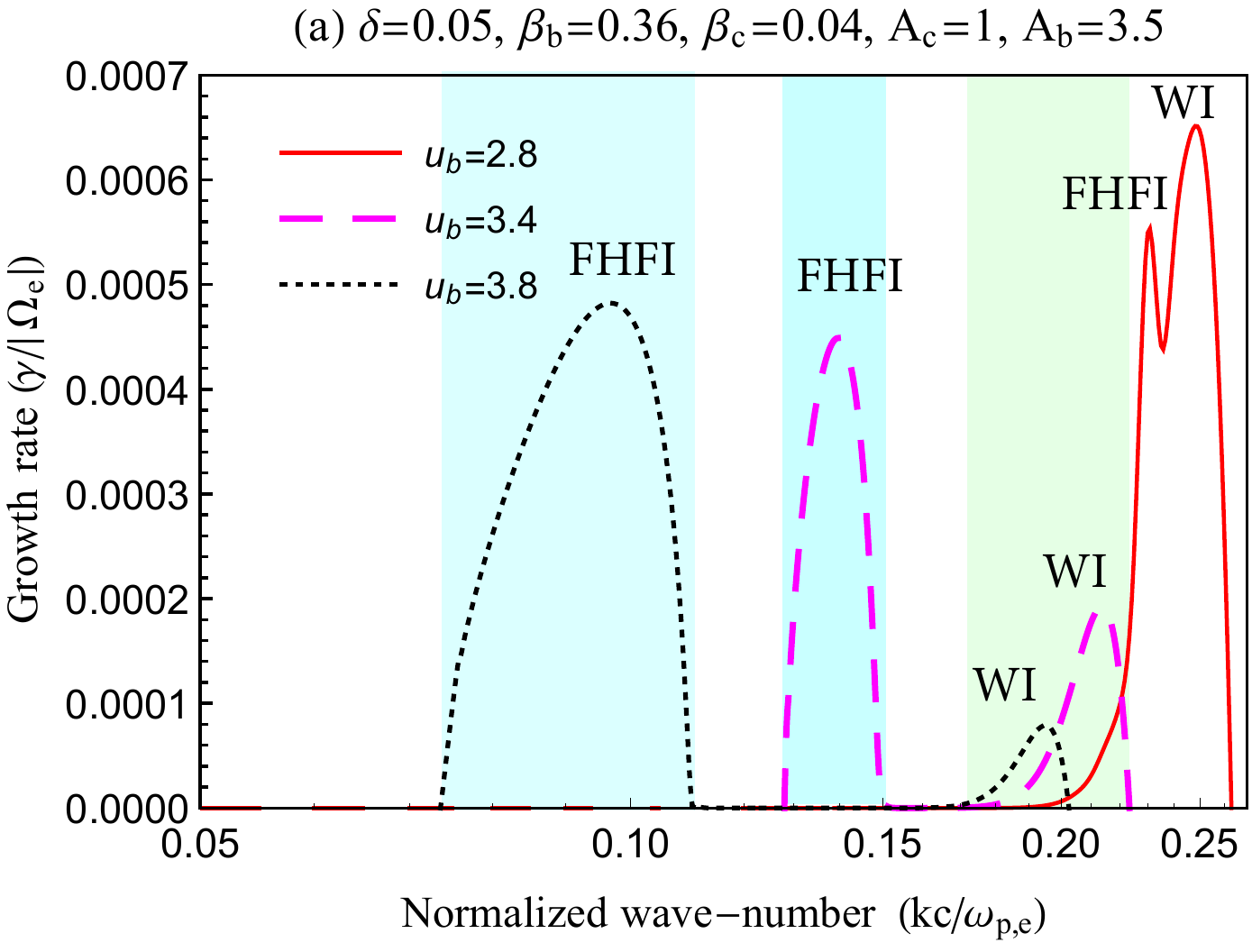} ~~~ \includegraphics[width=16.5pc]{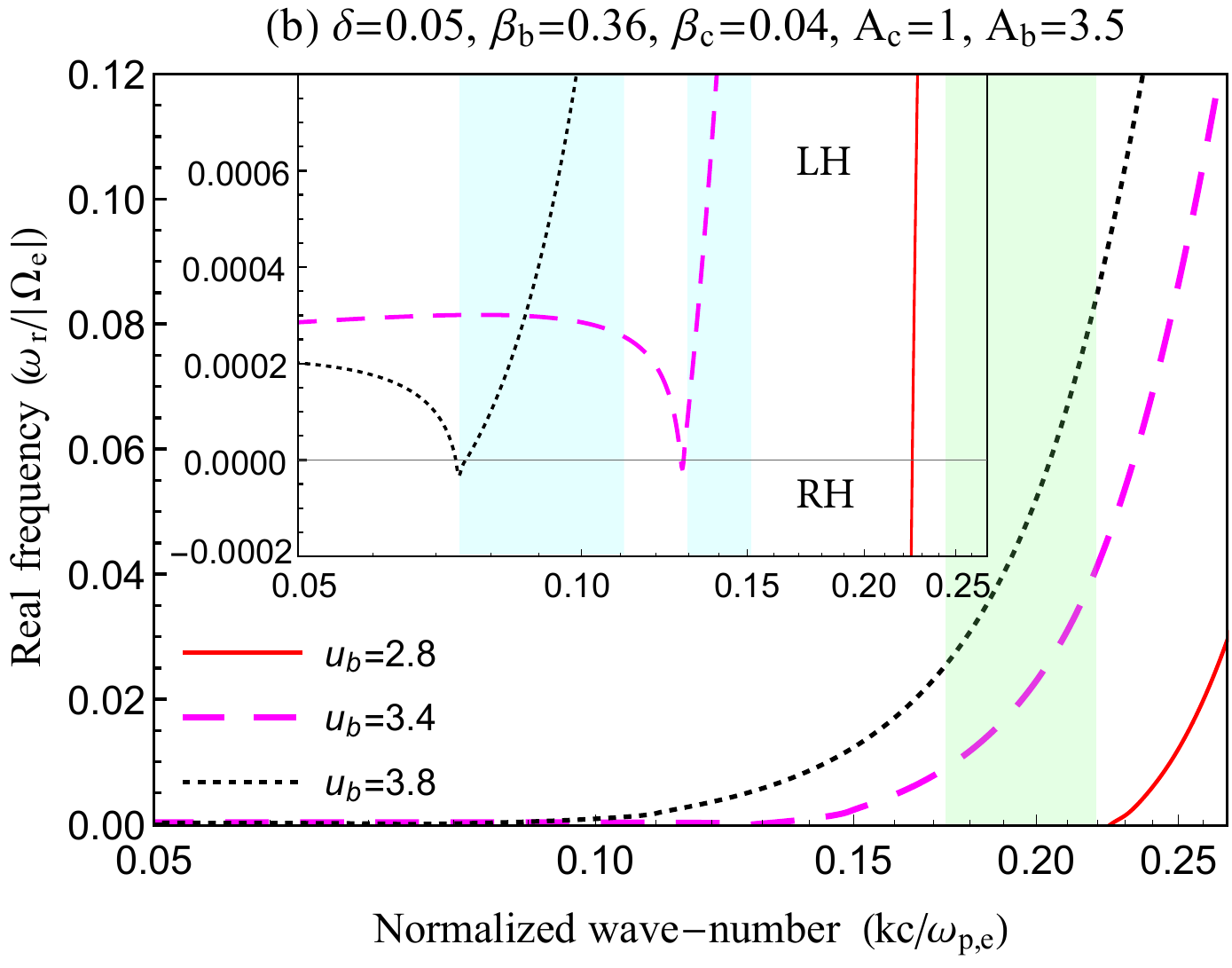}
\caption{Effect of the beam velocity $u_b$ on the growth rates (panel (a)) and wave frequency 
(panel b) of the WI instability driven by beam anisotropy $A_b=3.5$.} \label{f7}
\end{figure*}
%

Figure~\ref{f4} presents four distinct regimes of destabilized whistler modes, assuming $\delta = 0.05$, 
$A_c = 1.0$ and $\beta_c = \beta_p = 0.04$. In panel~(a) we consider, for reference, the beam isotropic 
$A_b=1.0$, and find growth or damping rates of whistlers varying only under the influence of the 
beam velocity $u_b$: the instability is obtained for higher beaming velocities $u_b = 0.5$, 0.6, 
and maximum growth rates are obtained for $u_b=0.6$ \citep{Shaaban2018}. In panel~(b), for a 
relatively small beam anisotropy $A_b=1.2$, growth rates are markedly stimulated by the beam velocity, 
and maximum growth rates are obtained for a less energetic beam, i.e., $u_b=0.5$ (brown line). Higher 
anisotropies $A_b=1.5, 2.5$ may drive an instability with significantly high growth rates (even in the 
absence of a beam, $u_b=0$), see red lines in panels (c) and (d). In panel (c) the instability features 
characteristics of both the WHFI and WI, cumulating the effects of beam and temperature anisotropy. 
Maximum growth rates are obtained for $u_b=0.3$. For higher anisotropies $A_b=2.5$, in panel (d), we obtain 
WI-like growth rates (maximum for $u_b =0$) which decrease as the beam velocity increases. More energetic 
beams, e.g., $u_b=0.6$ (black line), may determine another transition to WHFI regime. 

A series of conclusions can already be drawn, which enable to distinguish between these two regimes of unstable 
whistlers (also see next section of the instability thresholds). Thus, the beam anisotropy $A_b>1.0$ 
stimulates the WHFI, reducing also the beam velocity required for the instability to display maximum 
growth rate. On the other hand, growth rates of WI are reduced by the beam, and an increase of $u_b$ 
may trigger a transition to the WHFI.

\section{Unstable electron firehose modes}\label{sec.4}

In this section, we investigate the LH branch of HFIs represented by the electron firehose heat-flux 
instability (FHFI). Conditions of this instability are expected to be markedly modified under the influence
of temperature anisotropies $A_{b,c} < 1$, which are responsible for the excitation of standard firehose 
instability (FI). For sufficiently large core plasma beta $\beta_c$ and high beaming velocity $u_b>2.7$ 
both the FHFI and FI are expected to develop with similar dispersive features \citep{Gary1993, Shaaban2018}. 
First we analyze the FHFI under the mutual effects of the electron beam ($u_b \ne 0$) and its temperature 
anisotropy ($A_b \neq 1$). The unstable solutions are obtained by solving numerically the dispersion relation 
(\ref{eq:dis}) for LH modes with $\xi_p^-$. By contrast to recent studies of FHFI, which consider only 
small plasma beta regimes, i.e., $\beta_c=0.04$ \citep{Saeed2017b}, here we assume solar wind high beta 
conditions, i.e., $\beta_c= \beta_p > 1$, which are more favorable to FHIs. 

In Figures~\ref{f5} and \ref{f6} we assume $\delta=0.05$, $\beta_c=~\beta_p=~1.2$, $A_c=1.0$, and more 
energetic beams $u_b=3.8$. Figure~\ref{f5} shows the effects of an increasing anisotropy $A_{b}=1.0, 
1.2, 1.4$ on the FHFI: in panel (a) growth rates decrease and the range of unstable wave numbers increases, 
and in panel (b) the wave frequency exhibits the same monotonous increasing. An opposite anisotropy $A_b=~1.0, 
0.7, 0.5$, assumed in Figure~\ref{f6} has a cumulative effect stimulating the FHFI by increasing the growth 
rates and wave-frequencies. The wave frequency keeps the positive sign $\omega_r>0$ in the range of the FHF 
peaks. The core anisotropy $A_c \neq 1$ manifests similar effects on the FHFI (not shown here).

In the previous section we have outlined a transition from WI to WHFI, triggered by the increase of 
the beam speed $u_b$, when the temperature anisotropy of the beam is relatively small. Here in 
Figure~\ref{f7}, we show that, provided the anisotropy is high enough, i.e., $A_b = 3.5$, WI can 
directly convert to FHFI with increasing $u_b = 1.2, 2.0, 2.8$. The WI is driven by the beam 
anisotropy $A_b=3.5$ for the same plasma parameters invoked in \cite{Saeed2017b} (their Figure~3):
$\beta_c=0.04$, $\beta_b=0.36$, $\delta=0.05$ and $ A_c=1.0$. Top panels present the first regime 
where the WI instability is dominant and the beaming velocity $u_b<2.8$ is below but close to the 
threshold value for the excitation of FHFI \citep{Saeed2017b, Shaaban2018}. Increasing the beam 
velocity has an inhibiting effect leading to a decrease of both the growth rates and the range of 
unstable wave numbers of WI (panel a). The corresponding wave frequencies (panel b) decrease and 
remain RH polarized ($\omega_r<0$) in the range of the WI instability peaks, unless for energetic 
beams when the polarization changes to LH (cyan areas) under the influence of FHFI which exhibit a 
second distinct peak of growth rates (red solid line). The double-peak growth rates is relevant for 
the transition between the two regimes of FHFI and WI. FHFI becomes dominant for more energetic
beams ($u_b>2.8$), when the FHF peak is markedly enhanced moving towards lower wave-numbers (bottom panels 
c and d). Small peaks of WI are decoupled and still visible, but remain LH polarized (green area).
Such reversals of the whistler mode polarizations have been observed by STEREO in the Earth's inner 
plasma-sphere at $L<2$  \citep{Breneman2011}. 

\begin{figure}[t]
\centering 
\includegraphics[width=18.5pc]{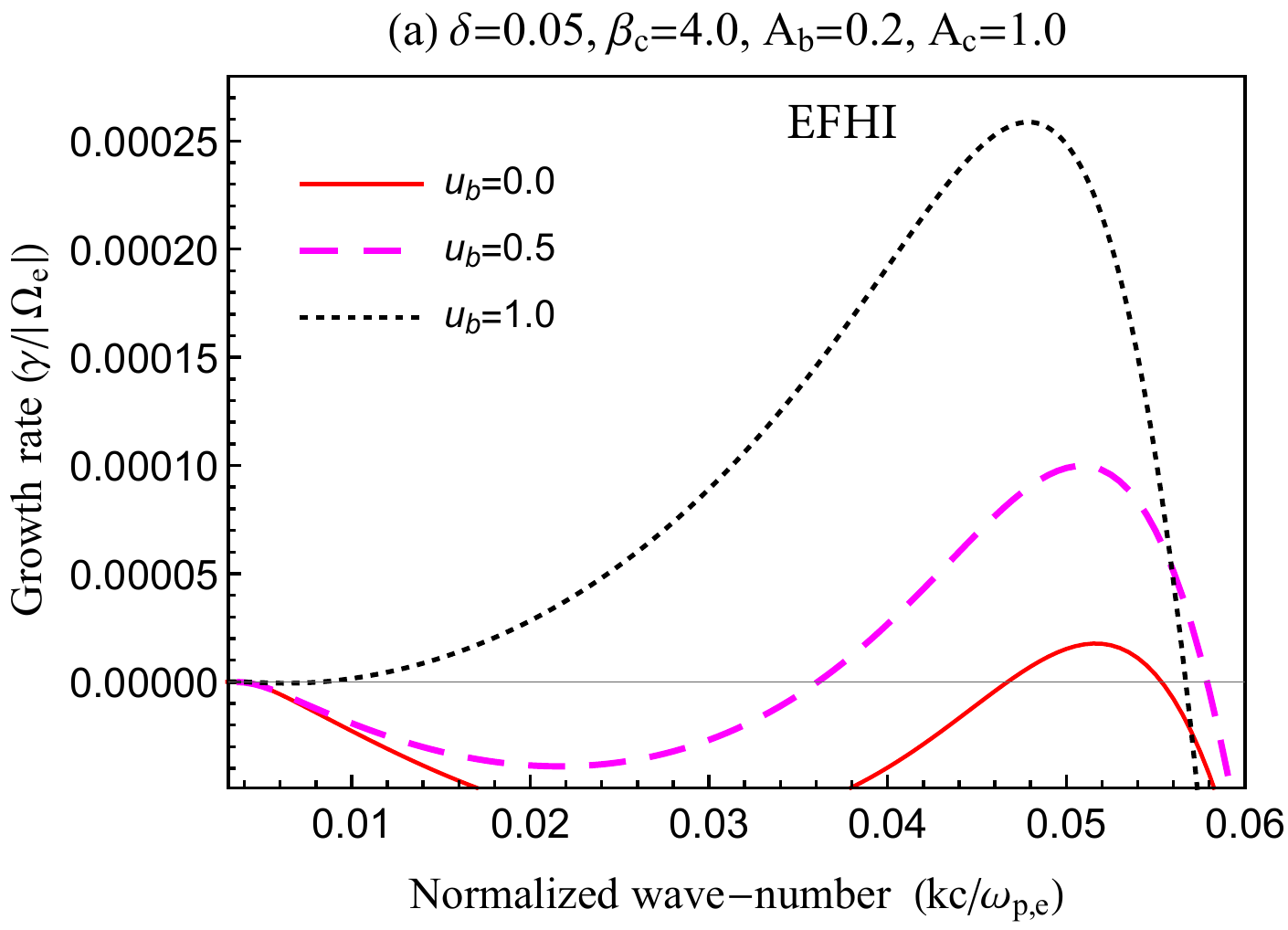} 
\includegraphics[width=18.5pc]{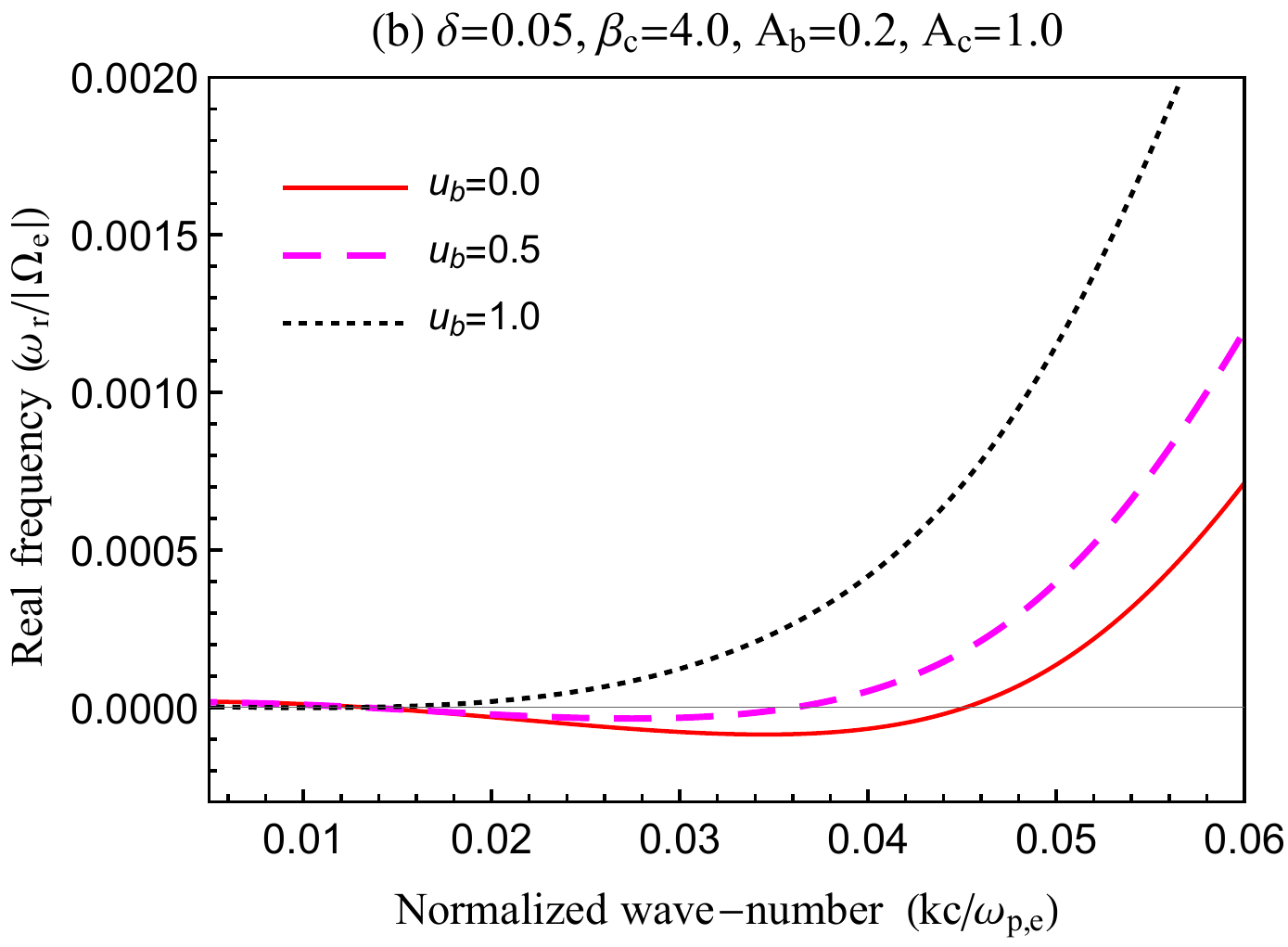}
\caption{Effect of the beam velocity $u_b$ on the growth rates (panel a) and wave frequency 
(panel b) of the EFHI driven by beam anisotropy $A_b=0.2$.}
\label{f8}
\end{figure}

For a core with a sufficiently large plasma beta, a beam with an excess of parallel temperature, i.e.,
$A_b < \, 1$ may excite the electron firehose (EFH) instability. Figure~\ref{f8} shows the effect of the 
beam velocity $u_b$ on the EFH instability driven by a temperature anisotropy $A_b=0.2$ for the 
following plasma parameters $\delta=0.05$, $\beta_c=\, \beta_p=\, \beta_b/10 = 4.0$, $A_c=1.0$. 
The growth rates (panel a) and wave frequencies (panel b) of EFH instability are significantly 
stimulated by increasing $u_b = 0.0, 0.5, 1.0$. These effects contrast with the inhibition
of WI shown in Figure~\ref{f7}.

%
\section{Threshold conditions} \label{sec.TC}

\begin{figure}[t]
\centering 
\includegraphics[width=17pc]{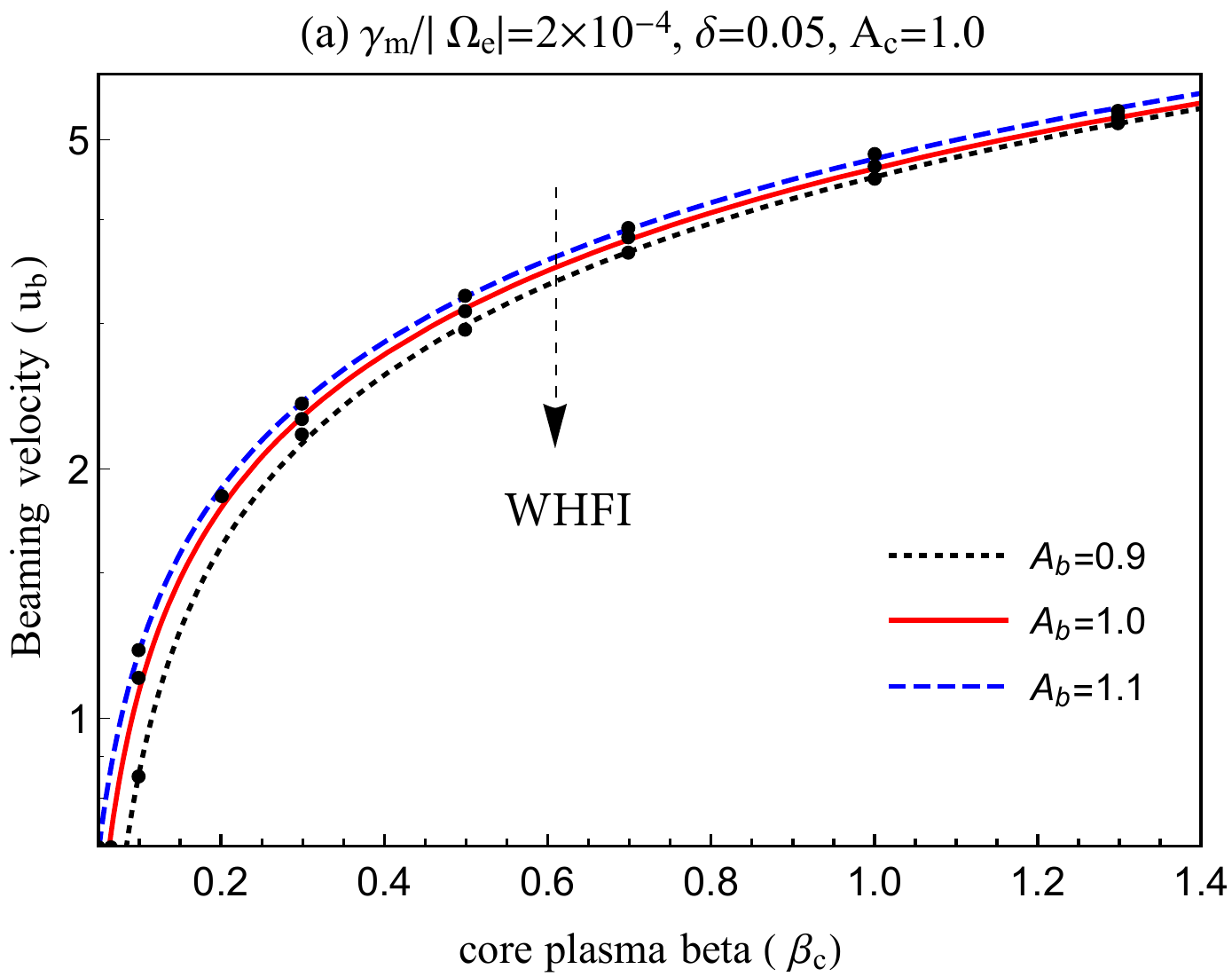} 
\includegraphics[width=17pc]{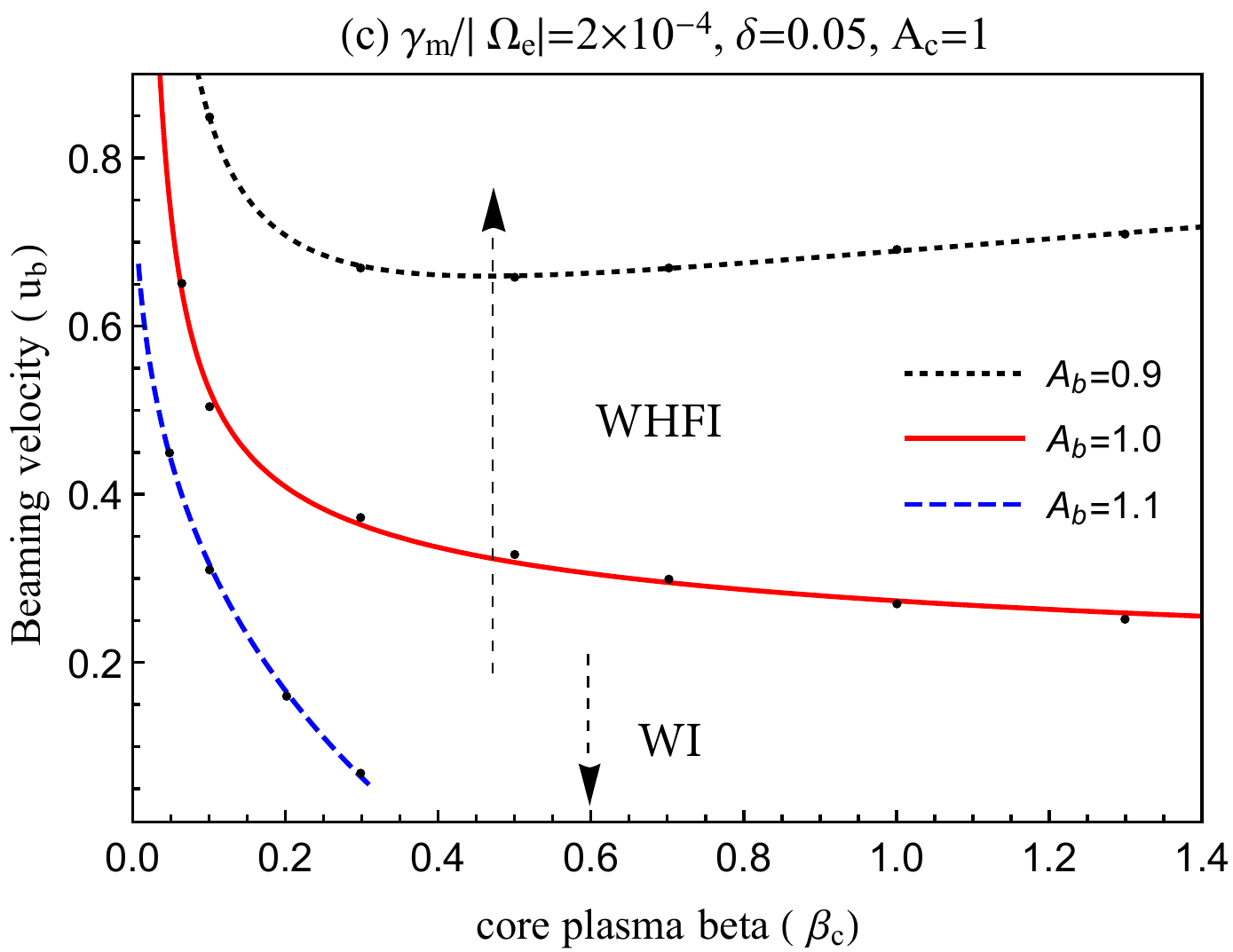}
\caption{Effect of the beam anisotropy $A_b$ on the (a) upper and (b) lower thresholds ($\gamma_m=2\times10^{-4} |\Omega_e|$) 
of the WHFI.}
\label{f9}
\end{figure}

\begin{figure}[t]
\centering 
\includegraphics[width=17pc]{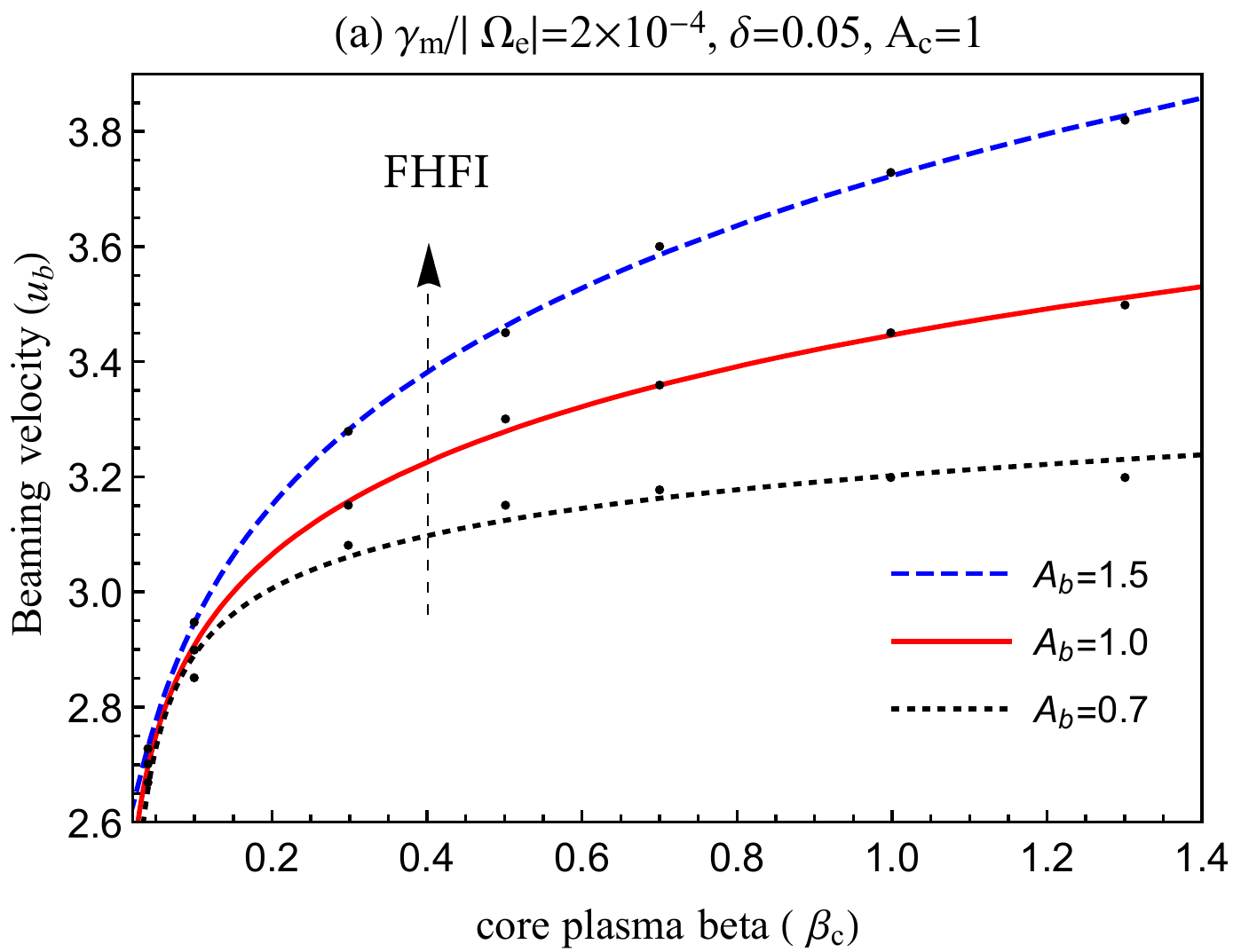} 
\includegraphics[width=17pc]{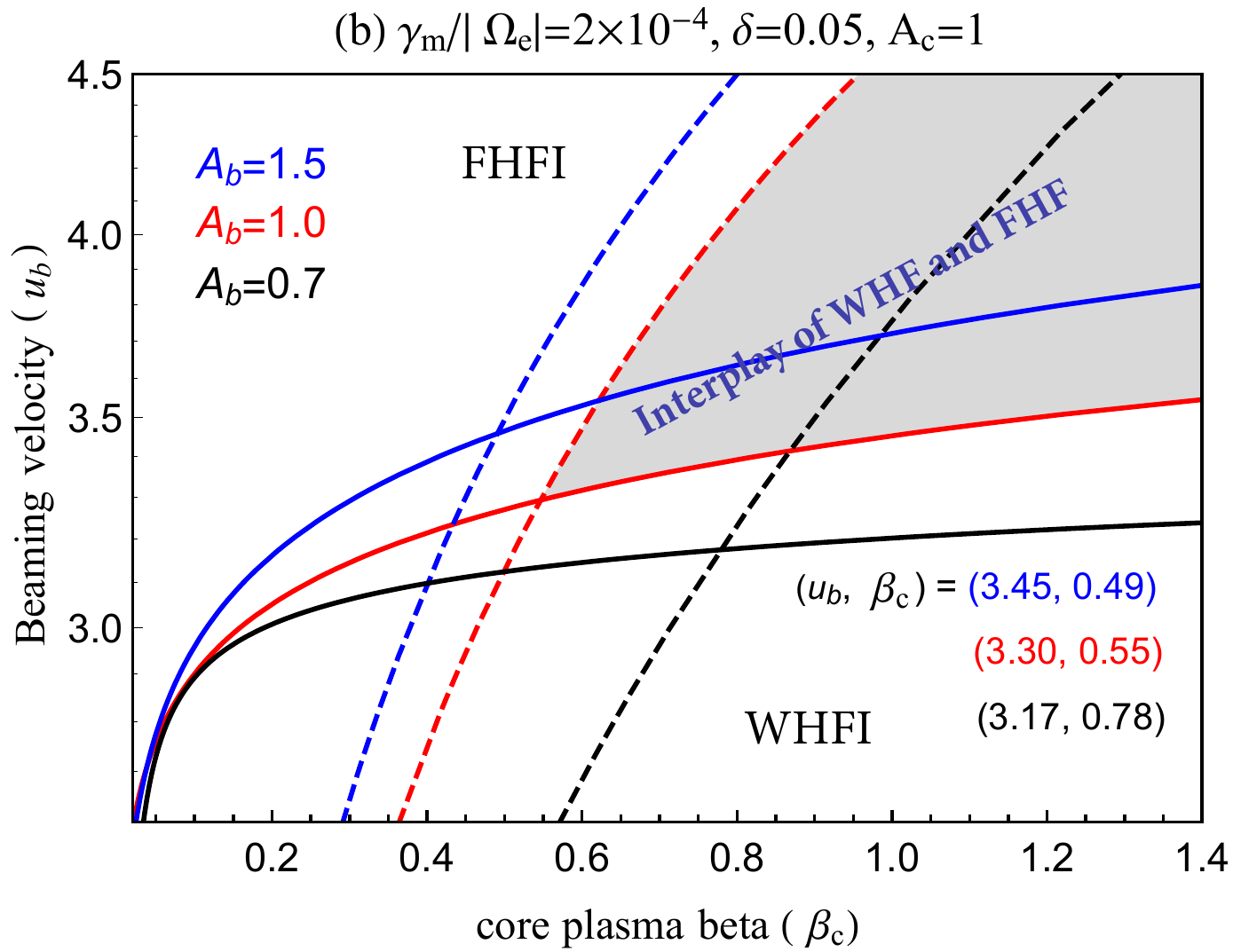}
\caption{Effect of the beam anisotropy $A_b$ on the (a) FHFI threshold and a comparison with the (b) WHFI upper threshold 
for $\gamma_m=~2\times10^{-4} |\Omega_e|$.}
\label{f10}
\end{figure}

\begin{figure}[t]
\centering 
\includegraphics[width=17pc]{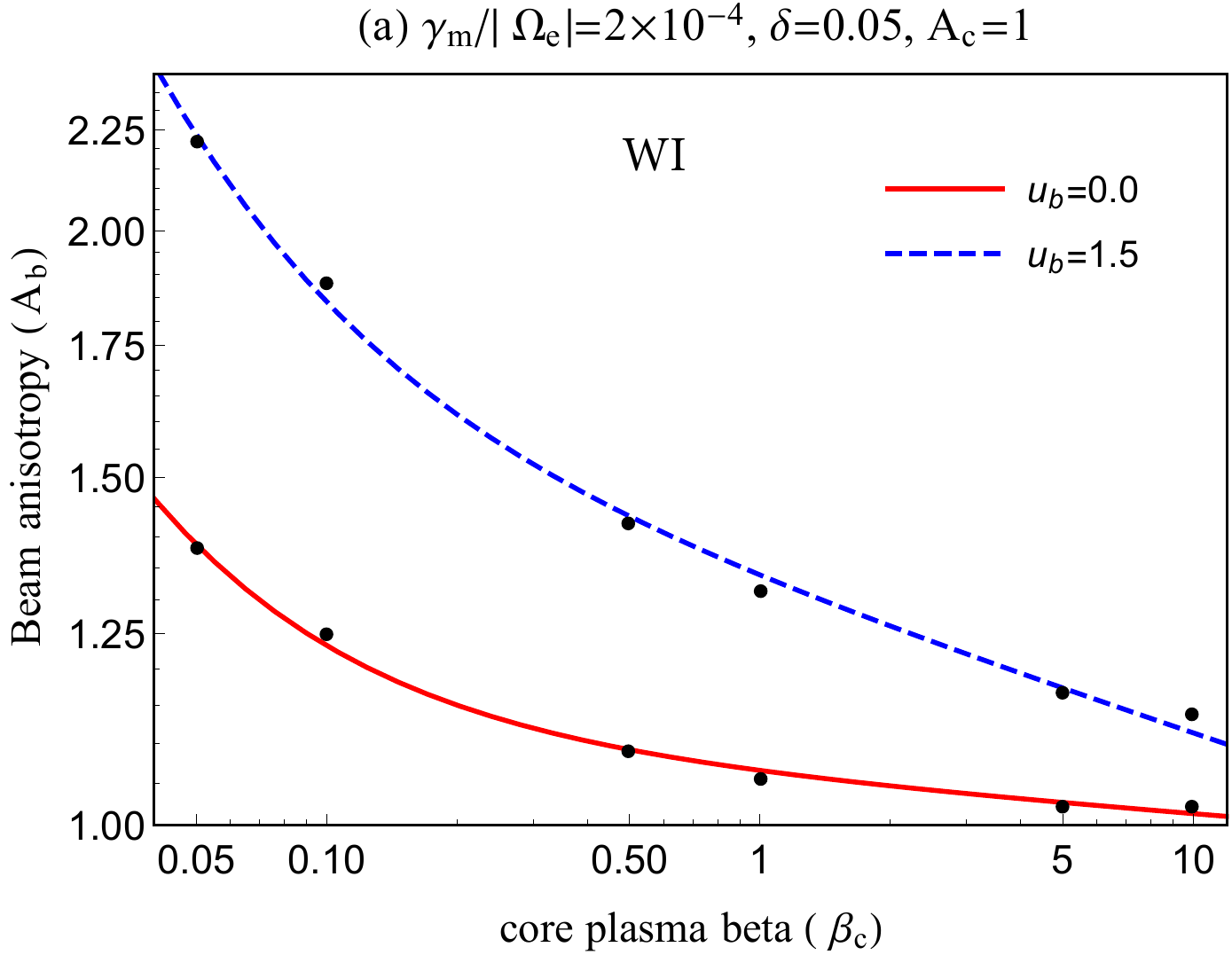} 
\includegraphics[width=17pc]{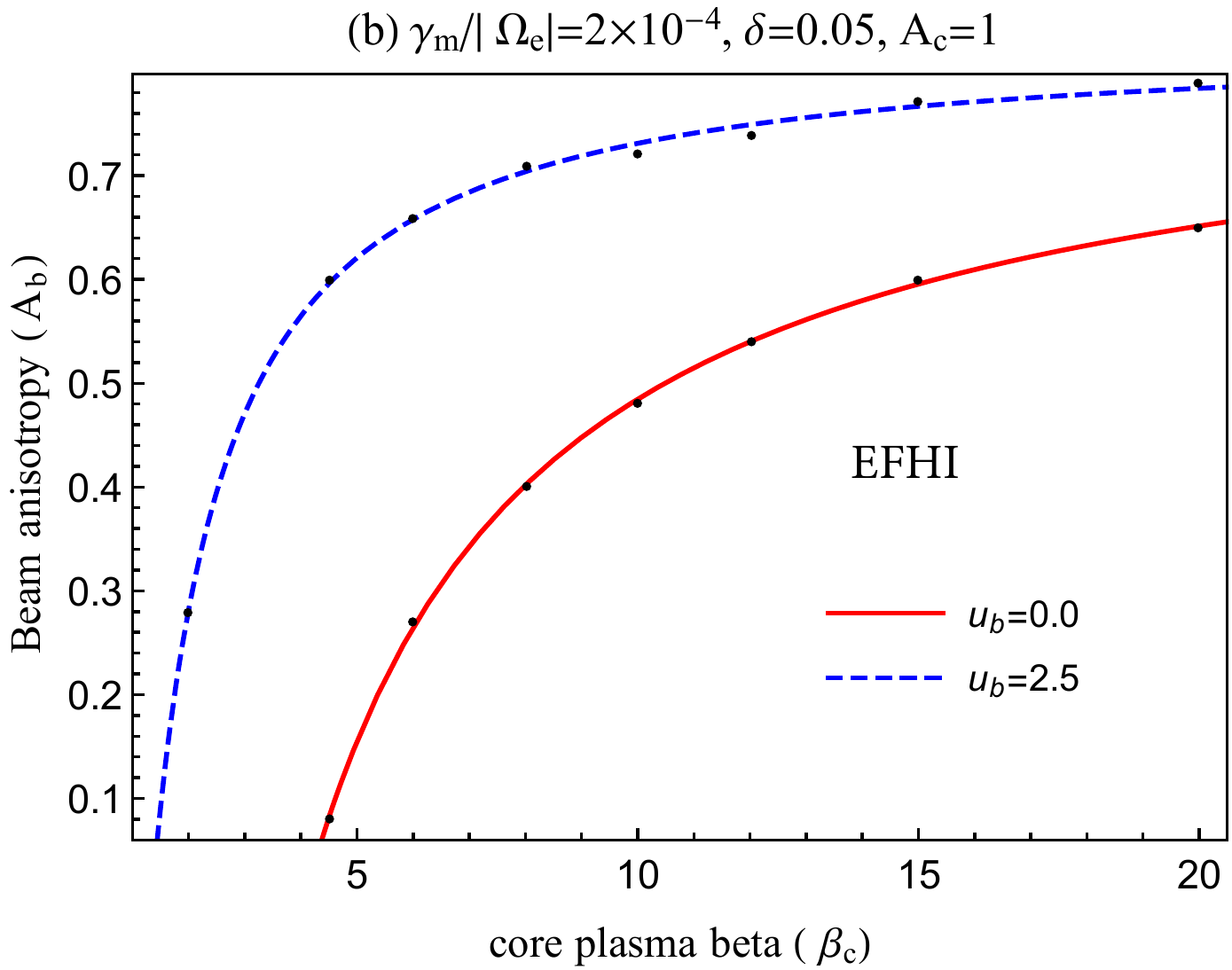}
\caption{Effect of the beam velocity $u_b$ on WI (panel a) and EFH (panel b) instabilities thresholds with maximum growth rates $\gamma=2\times10^{-4} |\Omega_e|$.  The plasma parameters are mentioned in each panel.}
\label{f11}
\end{figure}

Thresholds offer a concise but more comprehensive picture of the unstable regimes. Figures~\ref{f9}--\ref{f11} 
present the instability thresholds derived for a small maximum growth rate $\gamma_m= 2\times10^{-4}|\Omega_e|$, 
approaching marginal stability, i.e. $\gamma_m\rightarrow0$. These thresholds are derived in terms of the 
instability drivers, i.e., beam velocity $u_b$ or temperature anisotropy $A_b$, as a function of the core 
plasma beta $\beta_c$. The other plasma parameters are kept constant, e.g., $\delta=0.05$ and $A_c=1.0$. 
Mathematically, the instability thresholds are fitted to a function of $\beta_c$ genericalle expressed by 
\citep{Shaaban2016}
\begin{align}
\Delta=\left(1+\frac{a}{\beta_c^ {~b}}\right)\frac{c}{\beta_c^ {~d}}\label{TC}
\end{align}
where
\[
     \Delta= 
\begin{cases}
    A_b,& \text{for } \text{temperature anisotropy instabilities}\\
    u_b,& \text{for } \text{heat flux instabilities. }
\end{cases}
\]
Fitting parameters $a$, $b$, and $c$ are tabulated in Tables~\ref{t2}-\ref{t4} in Appendix.
%
%

%
%
As shown in Figure~\ref{f4}~(a), the WHFI growth rates vary non-uniformly with increasing the beaming velocity, 
suggesting, as also shown recently by \cite{Shaaban2018}, that the unstable WHF modes are bounded between two 
thresholds of the beam velocity. Figure~\ref{f9} describes the effect of temperature anisotropy $A_b\neq1.0$ 
on the upper and lower thresholds of WHFI, in panels (a) and (b), respectively. We contrast thresholds for
an isotropic beam ($A_b=1$, solid-red), with those for $A_b=1.1$ (dashed-blue) and $A_b=0.9$ (dotted-black). 
In panel (a) the upper threshold is slightly increased by the anisotropy in perpendicular direction, $A_b=1.1$, 
but it is slightly decreased by an opposite anisotropy in parallel direction, $A_b=0.9$. Only small variations 
are obtained in this case, given that this regime of WHFI is mainly controlled by relatively high beaming 
velocities, and temperature anisotropies are relatively small. However, for the same anisotropies, in panel (b) 
the lower WHFI threshold undergoes more important changes. This threshold is markedly enhanced for $A_b=0.9$, 
and this difference is increased with increasing the core plasma beta $\beta_c$, squeezing  the unstable regime 
of WHFI (upper directed arrow). In an opposite situation for $A_b=1.1$, the lower WHFI threshold is markedly 
reduced to lower beaming velocities $u_b$, and with increasing $\beta_c$ the WI peak starts to dominate the WHFI peak, which finally 
quenches completely. This regime marks the transition from WHFI to the most common WI (lower directed arrow) 
which becomes exclusively driven by the temperature anisotropy (for $u_b = 0$ and $\beta > \beta_c \simeq 0.3$).
 
Figure~\ref{f10}~(a) describes the effect of temperature anisotropy on the FHFI thresholds, by contrasting 
thresholds derived for an isotropic beam ($A_b=1$, solid-red) with the those for $A_b>1.0$ (dashed-blue) and 
$A_b<1.0$ (dotted-black). Variations in this case resemble those of the WHFI upper threshold, but the most unstable 
FHF modes are situated above the thresholds, as pointed out by the dashed arrow. Temperature anisotropy $A_p=1.5$ 
has a stimulating effect on the FHFI threshold, squeezing the unstable regime of the FHF modes, while for $A_b=0.7$ 
the threshold is reduced and the FHFI regime is enlarged. These effects are boosted by increasing $\beta_c$, 
confirming the results in Figures~\ref{f5} and \ref{f6}. 
For the sake of comparison, in panel~(b) we compare the FHFI and the WHFI thresholds for the same 
set of plasma parameters ($\delta=0.05$, $A_c=1.0$), and different anisotropies $A_b=0.7, 1.0, 1.5$. As shown in 
panel~(a), the unstable FHF modes require relatively high beaming velocities $u_b>2.7$ making the comparison 
relevant only for the WHFI upper threshold. For isotropic beams (red lines) the FHFI is dominant at low 
$\beta_c<0.55$ and high $u_b>2.7$, and also at beaming velocities exceeding the WHFI (upper) threshold, 
while the WHFI is dominant for less energetic beams with $u_b<2.7$. If $u_b>3.3$ and $\beta_c>0.55$ are high enough, 
we can identify a regime of transition (gray shaded regime) where the unstable FHF and WHF modes may co-exist and interplay. 
The anisotropic beams with $A_b=1.5$ determine the interplay regime to move towards higher $u_b>3.45$ and lower 
limit $\beta_c>0.49$. For an opposite anisotropy $A_b>0.7$ this regime moves towards lower $u_b>3.17$ and higher 
limit $\beta_c>0.78$.
 
In order to complete the analysis, in Figure~\ref{f11} we describe the effect of the beam velocity on the thresholds 
of temperature anisotropy driven instabilities, WI in panel (a) and FI in panel (b). As expected, the WI threshold is markedly 
enhanced by the beam velocity $u_b=1.5$, confirming the inhibiting effect already shown on the growth-rates in Figures~\ref{f3} 
and \ref{f7}. Here we can see that this effect is reduced with increasing $\beta_c$. Also expected is the effect 
shown by the EFH threshold, which is markedly reduced in the presence of beam, see panel~(b), confirming the stimulating 
effect of beams on the EFHI growth rates, obtained in Figure~\ref{f8}.
 
\section{Discussions and conclusions}\label{sec.6}   

As shown in the introduction, the heat-flux instabilities may play a major role in the evolution of electron beams
in the solar wind, but a definitive answer on this issues requires a detailed examination of these instabilities
in conditions specific to space plasmas. The kinetic approach proposed in this paper enables an advanced 
characterization of the heat-flux instabilities for complex but realistic conditions, when the electron beams 
exhibit temperature anisotropies. The new unstable regimes uncovered here are controlled by two drivers, i.e., 
beaming velocity $u_b$ and beam anisotropy $A_b$, and by the core plasma beta $\beta_c$, and we have contrasted
with idealized regimes of instabilities driven either by isotropic beams \citep{Saeed2017, Shaaban2018} or 
by non-drifting (core-beam) populations with temperature anisotropies.

For less energetic beams the WHFI is found to be very sensitive to the beam anisotropy: growth rates are markedly 
increased if $A_b>1.0$, and are decreased when $A_b<1.0$ (Figure~\ref{f1}). Core anisotropy $A_c\neq 1.0$ shows similar 
effects on the WHFI, but it is much less effective than the beam anisotropy, see Figure~\ref{f2}. The common WI can be 
excited at low $\beta_c < 1$ but for a significant $A_b > 1.0$. The beam has an inhibiting effect on the WI: reducing  
growth rates and the range of unstable wave-numbers with increasing the beaming velocity. But, apparently, the beam may 
stimulate WI driven by the core anisotropy is $A_c>1.0$, see Figure.~\ref{f3}, where growth rates increase and saturate 
for higher beaming velocities, resembling a regime characteristic to WHFI. 

Firehose instabilities are expected to develop for relatively higher $u_b$ or/and higher $\beta_c$, and contrary to 
whistlers, differences between FHFI and FI are easier to determine. FHFI can be excited even for a low $\beta_c < 1$, 
provided the beam velocity is high enough, while FI requires a high $\beta_c\geqslant3.0$ and a temperature anisotropy 
$A_b<1.0$. For a moderately high $\beta_c=1.2$, we have found that $A_b <1$ has a stimulating effect on the FHFI, 
increasing the growth rates and the corresponding wave frequencies (Figure~\ref{f5}). The anisotropy in perpendicular 
direction has an opposite effect, see Figure~\ref{f6}, where growth rates and wave frequencies of FHFI decrease with 
increasing $A_b > 1$. These variations of the growth rates and wave frequencies with the temperature anisotropies have 
not been observed in the previous studies which were restricted to low $\beta_c=0.04$ regimes \citep{Saeed2017b}.

Figure~\ref{f7} suggests that, depending on the beaming velocity $u_b$, the interplay with temperature anisotropy 
$A_b>1.0$ can be divided into two distinct regimes. For a beaming velocity below the threshold of FHFI, i.e., $u_b<2.8$, 
dominant is the WI, and, as expected, the beam has an inhibiting effect, reducing the growth rates and the range of 
unstable wave-numbers. 
In the second regime, more energetic beams with $u_b\geqslant2.8$ excite the FHFI and the growth rates display a second 
distinct peak at low wave-numbers. The beam stimulates the FHFI, but inhibits the WI peak, and wave frequencies 
may change sign showing a LH polarization even in the range of the WI peak, under the influence of a dominant FHFI. 
Regarding the more common FHI driven by an excess of parallel temperature ($A <1$), the effective free energy is 
enhanced in the presence of a beam, and FHI is stimulated (Figure~\ref{f8}). 
Another remark can be made if we calculate the core drift velocity $u_c$ for the same plasma parameters used to derive 
the heat-flux unstable modes in Figures~\ref{f4}~(a) and \ref{f7}~(b). The plasma parameters used for the WHFI, e.g.,
$\delta=0.05$, $\beta_c=0.04$, and $u_b=U_b/c ~\omega_e/|\Omega_e|=~0.6$, where $\omega_e/|\Omega_e|=100$, 
imply for the core drift velocity $U_c=~\delta~U_b/(1-~\delta)=~3.16\times 10^{-4}$, which is about 1.6 times higher 
than Alfv\'en velocity $V_A=~2\times~10^{-4} c$,  commonly invoked in similar studies. For the FHFI we assumed 
$\delta=0.05$, $\beta_c=0.04$ and $u_b=3.8$, implying a higher core drift velocity $U_c=10~V_A$. In the solar 
wind $U_c$ is comparable to, or larger (three times larger in a collisionless plasma) than $V_A$ \citep{Pulupa2014}. 
Thus, our results strengthen the early predictions \citep{Gary1975, Gary2000} that whistler instabilities could be 
more efficient in regulating the electron heat flux in the solar wind.

Thresholds displayed in Figures~\ref{f9}--\ref{f11} may provide a better overview on the interplay of these instabilities.
In Fig.~\ref{f9} the unstable WHF modes are located between two thresholds, namely, a lower and an upper threshold. 
In terms of the beam velocity $u_b$, the interval of WHFI in between these two thresholds may significantly increase
even for a modest temperature anisotropy in perpendicular direction $A_b>1$, or it is markedly reduced by a 
an opposite anisotropy in parallel direction $A_b<1$. Situated above these thresholds, the unstable regime of the 
FHF modes, see Figure~\ref{f10}~(a), show opposite effects, increasing when $A_b < 1$ and diminishing for $A_b > 1$. 
All these variations increase with increasing $\beta_c$. In Figure~\ref{f10}~(b), we have identified unstable regimes
conditioned by both the WHFI and FHFI, which move either towards higher $u_b$ and lower $\beta_c$ if the anisotropy 
increases in perpendicular direction ($A_b>1$), or towards lower $u_b$ and higher $\beta_c$ by increasing the 
anisotropy in parallel direction ($A_b<1.0$). These unstable regimes are considerably enhanced by increasing 
$\beta_c$. In Figure~\ref{f11} we have described the effects of beam on the temperature anisotropy thresholds. 
The WI threshold is increased by increasing the beaming velocity, confirming the inhibiting effect on the 
growth rates in Figures~\ref{f3} and \ref{f7}. On the other hand, the FI threshold is decreased by increasing the 
beaming velocity, confirming the stimulating effect on the growth rates in Figure~\ref{f8}. 

To conclude, we have identified new regimes of the whistler and firehose unstable modes, 
which are highly conditioned by the interplay of two sources of free energy, an electron 
beam and its intrinsic temperature anisotropy. Present study is focused on parallel electromagnetic 
modes, with intention to facilitate the analysis and differentiate between different regimes of 
these instabilities. In the oblique directions very efficient may be the aperiodic 
instabilities, like electron mirror or electron firehose, but their properties are known only for 
regimes triggered by the temperature anisotropies \citep{Maneva2016, Shaaban2018b}. Our results 
should therefore stimulate further investigations to address the full spectrum of beam-driven 
electromagnetic and electrostatic instabilities.

\acknowledgments
\begin{footnotesize}{The authors acknowledge support from the Katholieke Universiteit Leuven, Ruhr-University Bochum. 
These results were obtained in the framework of the projects 
SCHL 201/35-1 (DFG-German Research Foundation), GOA/2015-014 (KU Leuven), G0A2316N (FWO-Vlaanderen), 
and C 90347 (ESA Prodex 9). S.M. Shaaban would like to acknowledge the support by a Postdoctoral 
Fellowship (Grant No. 12Z6218N) of the Research Foundation Flanders (FWO-Belgium) and the support by a Travel Grant for a long stay aboard  
(Grant No. V419818N) of FWO-Belgium. P.H.Y. acknowledges NSF grant AGS1550566 to the University of Maryland, the Science Award Grant from the GFT Charity, Inc., to the University of Maryland, and the BK21 plus program from the National Research Foundation (NRF), Korea, to Kyung Hee University.
Some basic ideas developed in this paper have been discussed at the 1st ISSI meeting of the international team: Kappa Distributions.}\end{footnotesize}

\begin{appendix}
\section{Fitting parameters in Eq.~(\ref{TC})}

We have used Eq.~(\ref{TC}) to describe the instability plasma conditions in terms of the instability 
thresholds from Figures~\ref{f9}--\ref{f11}, defined by either the beam velocity or the temperature 
anisotropy (two distinct drivers), as a function of the core plasma beta. The fitting parameters $a$, 
$b$, $c$, and $d$ are tabulated in Tables \ref{t2} and \ref{t3} for the heat flux instabilities (WHFI 
and FHFI), and in Table \ref{t4} for the anisotropy driven instabilities (WI and FI).
%
\begin{table}[h]
\centering
\caption{Fitting parameters in Figure~\ref{f9}}
\label{t2}
\begin{tabular}{ccccccccccccccccccc}
\hline 
 & \multicolumn{3}{c}{WHFI (a)}&& \multicolumn{3}{c}{WHFI (b)}\\
 $A_b$ & 0.9        &  1.0   & 1.1        &&      0.9     &   1.0      &  1.1\\
 \hline 
  $a$    &  0.13      & 0.03 & $-$1.32  && $-$0.023 & $-$0.04 & $-$0.01  \\ 
 $b$     &  1.0        & 1.0   & $-$0.29  &&  1.0         & 1.0        & 1.0  & \\
 $c$     & 0.30      & 0.26 & 0.0         &&  4.72        & 4.67     &  4.81\\
 $d$     & $-0.22$ & 0.18 & 1.0      &&  $-$0.52  &$-$0.54 & $-$0.53\\

 \hline 
\end{tabular}
\end{table}
%
\begin{table}[h]
\centering
\caption{Fitting parameters in Figure~\ref{f10}}
\label{t3}
\begin{tabular}{ccccccccccccccccccc}
\hline 
 & \multicolumn{3}{c}{FHFI} && \multicolumn{3}{c}{WHFI}\\
 $A_b$ & 0.7        &  1.0   & 1.5         && 0.7        &  1.0   & 1.5\\
 \hline 
  $a$    & $-$0.003      & 2.45 &2.72  && 5.82      &$-$0.023 & $-$0.008  \\ 
 $b$     & 1.0      &$-$ 0.1 & $-$0.14  && 1.0        & 1.0   & 1.0  \\
 $c$     & 3.21      & 0.0 & 0.0            &&0.55      & 4.72 & 5.11\\
 $d$     & $-$0.03     &1.0 & 1.0         &&$-$1.54 & $-$0.52 & $-$0.53\\

 \hline 
\end{tabular}
\end{table}
%
\begin{table}[h]
\centering
\caption{Fitting parameters in Figure~\ref{f11}}
\label{t4}
\begin{tabular}{ccccccccccccccccccc}
\hline 
 && \multicolumn{2}{c}{WI} && \multicolumn{2}{c}{FI}\\
 $u_b$ && 0.0        &  1.5          &&      0.0     &   2.5\\
 \hline 
  $a$ && 0.012     & 0.018   & & $-$4.0   & $-$1.34\\ 
 $b$    && 1.0        & 1.0            &&  1.0           & 1.0      \\
 $c$     &&1.05      & 1.32          &&  0.80         & 0.85      \\
 $d$  &&0.017  & 0.072        &&  $-$0.004  & 0.006 \\

 \hline 
\end{tabular}
\end{table}
%

\end{appendix}
\bibliographystyle {agsm}
\bibliography{papers}
	 
\end{document}